\documentclass[authoryear,preprint,12pt]{article}

\makeatletter
\def\ps@pprintTitle{%
 \let\@oddhead\@empty
 \let\@evenhead\@empty
 \def\@oddfoot{}%
 \let\@evenfoot\@oddfoot}
\makeatother

\usepackage[]{amsfonts}
\usepackage[]{amsmath}
\usepackage[]{amssymb}
\usepackage{amsthm}
\usepackage{anysize}
\allowdisplaybreaks
\usepackage{accents}
\usepackage{afterpage}
\usepackage{adjustbox}

\usepackage{bbm}
\usepackage{booktabs}

\usepackage{cancel}
\usepackage{capt-of}
\usepackage[bf]{caption}
\usepackage{color,colortbl}
\usepackage{calligra}
\RequirePackage[colorlinks=true,citecolor=blue,urlcolor=blue]{hyperref}

\usepackage{empheq}
\usepackage{enumitem}
\usepackage{epigraph}

\usepackage{fancyhdr}
\usepackage{float}

\usepackage{graphicx}

\usepackage{endnotes}
\let\footnote=\endnote
\usepackage{inputenc}

\usepackage[]{latexsym}

\usepackage{mathrsfs}
\usepackage{mathtools}
\mathtoolsset{showonlyrefs}
\usepackage{makecell}
\usepackage{multirow}

\usepackage{natbib}

\usepackage{tabularx}
\usepackage{tikz}

\usepackage{placeins}

\usepackage{setspace}
\usepackage{subcaption}

\usepackage{tabularx}

\usepackage{url}
\usepackage[normalem]{ulem}

\usepackage[left=1in,right=1in,top=1.5in,bottom=1.5in]{geometry}

\usepackage{lscape}
\usepackage{longtable}
\usepackage{pdfpages}
\usepackage{pdflscape}
\usepackage{rotating}
\usepackage{changepage}

\usepackage{multibib}
\newcites{supp}{Supplementary References}

\usepackage{natbib}
    \setcitestyle{aysep={}}

\usepackage{chngcntr}
\counterwithout{figure}{section}

\definecolor{Gray}{gray}{0.9}

\newcommand{\cR}{\mathcal{R}}

\newcommand{\bE}{\mathbb{E}}

\newcommand{\cN}{\mathcal{N}}

\newcommand{\Vlp}{\mathcal{V}^{\cR,LP}}

\newcommand{\bC}{\mathbb{C}}
\newcommand{\bR}{\mathbb{R}}

\newcommand{\bQ}{\mathbb{Q}}

\newcommand{\be}{\begin{equation}}
\newcommand{\ee}{\end{equation}}
\newcommand{\bew}{\begin{equation*}}
\newcommand{\eew}{\end{equation*}}

\newcommand{\dd}{{\rm d}}
\newcommand{\dt}{\dd t}
\newcommand{\dx}{\dd x}
\newcommand{\dZ}{\dd Z}
\newcommand{\dS}{\dd S}
\newcommand{\dV}{\dd V}
\newcommand{\dq}{\dd q}

\graphicspath{{../../../Figures/}{../../Figures/}{../../../../Figures/}{../../Figures/}}

\numberwithin{equation}{section}

\newtheorem{theo}{Theorem}[section]

\newtheorem{prop}[theo]{Proposition}

\makeatletter
\def\namedlabel#1#2{\begingroup
    #2%
    \def\@currentlabel{#2}%
    \phantomsection\label{#1}\endgroup
}
\makeatother

\definecolor{darkred}{RGB}{178,0,0}
\definecolor{darkgorange}{RGB}{220,88,42}
\definecolor{darkblue}{RGB}{0,0,139}
\definecolor{darkgreen}{RGB}{0,120,0}

\makeatletter
\def\blfootnote{\gdef\@thefnmark{}\@footnotetext}
\makeatother

\usepackage[deletedmarkup=xout]{changes}

\usepackage{todonotes}

\title{\large
\textbf{The Term Structure of Implied Fee Rates in Liquidity Pools}
}
\date{}
\author{Diogo Duarte\thanks{Florida International University, Adam Smith Center for Economic Freedom and College of Business, 11200 S.W. 8th St., 236, Miami, FL 33199, USA. Phone: +1-305-348-5429.  E-mail: \href{mailto:diogo.duarte@fiu.edu}{diogo.duarte@fiu.edu}.}
\quad
Lucas X. Ferreira\thanks{Rio de Janeiro, RJ, Brazil. E-mail: \href{mailto:lucasxistof@gmail.com}{lucasxistof@gmail.com}.}
\quad
Emmanuel Gobet\thanks{Laboratoire de Probabilit\'es, Statistique et Mod\'elisation, Sorbonne Universit\'e, 4 place Jussieu, 75005 Paris, France. E-mail: \href{emmanuel.gobet@sorbonne-universite.fr}{emmanuel.gobet@sorbonne-universite.fr}.}\ \thanks{Kaiko, 33 Rue du Louvre, 75002 Paris, France.}
\quad
Yuri F. Saporito\thanks{School of Applied Mathematics, Getulio Vargas Foundation (FGV), Brazil, Praia de Botafogo, Rio de Janeiro, RJ, Brazil. Phone: +55 21 3799-5917.  E-mail: \href{mailto:yuri.saporito@fgv.br}{yuri.saporito@fgv.br}.}
}

\begin{document}
\maketitle

\begin{abstract}
\noindent We develop a no-arbitrage framework to recover implied fee rates from option-implied information and quantify fee compensation for liquidity providers in automated market maker (AMM) pools. Expected accumulated fees equal the current pool value minus the risk-neutral value of its terminal payoff, reducing forward-looking LP compensation to pricing the LP payoff. We apply the framework to Uniswap v2 and Uniswap v3 ETH--USDC pools using a stochastic-volatility jump-diffusion model calibrated to Deribit options. 
Implied fee rates are positive, time varying, downward sloping in maturity, and highly sensitive to range design, with tighter near-the-money ranges earning larger implied returns. \\
\vspace{0.5in}

\noindent\textbf{JEL}: G20, G21
\vspace{0.15in}
\\
\noindent\textbf{Keywords}: Decentralized Finance, Liquidity Pools, AMM, Cryptocurrency
\vspace{0.15in}\\

\end{abstract}
\newpage

\onehalfspacing
\renewcommand{\baselinestretch}{1.1}







\section{Introduction}\label{sec:intro}

\paragraph{Context and problem.} Automated market makers (AMMs) have become a central market design in decentralized finance \citep{Adams2020,Adams2021}, to allow for on-chain exchanges of tokenized assets, with a similar purpose to a limit order book in traditional finance;
yet the compensation required to supply liquidity to these protocols remains difficult to measure ex ante. 
Once a liquidity provider (LP) has supplied her tokens in a concentrated-liquidity pool, she is exposed to a highly nonlinear payoff that depends on both the underlying asset price and the chosen liquidity range $\cR$ \citep{Adams2021,Neuder2021,Clark2021,Jin2022}. 
She is rewarded by earning fees, paid by swap traders for using the liquidity pool.
She collects these fees when she burns (i.e., withdraws) her initial supplied liquidity.

The LP's nonlinear payoff and the fees differ between Uniswap v2, where the liquidity range is infinite ($\cR:=(0,+\infty)$), and Uniswap v3, where the liquidity range is finite ($\cR:=[p_a,p_b)$ with $0<p_a<p_b<+\infty$), which allows for liquidity concentration.
Although fee mechanisms are transparent---with protocol code generally open-source and publicly auditable---quantitatively forecasting fee returns for a liquidity provider remains a difficult exercise.
The reason is that market participants typically do not observe a forward-looking measure of the compensation implied by prevailing market prices.
This creates a fundamental gap between AMM liquidity provision and the tools traditionally used in asset pricing and derivatives markets.

We bridge this gap by developing a framework to recover, from option-implied information, the \emph{implied total fee return} $R_{t,T}^{\cR}$ for an LP over any given time horizon $[t,T]$ when she has supplied liquidity on a range $\cR$ at time $t$. 
The key insight is that the value of a liquidity position can be decomposed into two components: the terminal value of the LP position within the pool and the fees accumulated over the holding period.
In the absence of arbitrage, the expected accumulated fees under a risk-neutral measure are equal to the difference between the current value of the LP position and the risk-neutral value of its payoff at time $T$. 
This effectively transforms the problem of measuring expected compensation for liquidity provision into the pricing of a vanilla claim whose payoff is the value at time $T$ of the LP position in the AMM.
Once this claim is priced using options data, the implied total fee return $R_{t,T}^{\cR}$ follows directly.

Leveraging the above methodology, our framework yields several new findings. 
First, the term structure $0\le\theta\mapsto \frac{1}{\theta}\log(1+R^\cR_{t,t+\theta})$ of continuously compounded rates is generally downward sloping: across calibration dates, longer holding horizons command lower implied fee rates.
At the same time, the level of the term structure varies substantially across dates, indicating that the compensation required to provide liquidity is highly state-dependent and moves with option-implied market conditions. 
Second, liquidity-range design plays a first-order role. 
In Uniswap v3, tighter ranges centered near the prevailing spot price generate systematically higher implied total fee returns than wider or more asymmetric ranges, reflecting the greater exposure of concentrated liquidity to local price fluctuations and inventory rebalancing.
Third, in Uniswap v3, the fee-rate surface is highly nonlinear in the location of the liquidity range and in maturity.
Together, these findings show that the expected profitability of liquidity provision depends jointly on horizon, market conditions, and the geometry of the liquidity band.
Lastly, we show that the pricing-based measure is economically meaningful when compared with realized fee generation. 
In backtests, model-implied expected accumulated fees track closely with realized fee accrual, even though the former is risk-neutral and forward-looking, whereas the latter is realized ex post.
This suggests that option-implied total fee returns capture an important component of realized LP profitability.
The time series of implied fee rates also exhibits pronounced variation across market states, with higher values during periods associated with more stressed or volatile market conditions. 
Finally, a principal-component decomposition of changes in fee-rate curves shows that the dynamics are highly low-dimensional: the first component explains 81.0\% of the variation, while the first three components jointly explain 99.6\%.

\paragraph{Related literature.}
The paper contributes to the growing literature on decentralized finance by linking AMM liquidity provision to option pricing and risk-neutral valuation.\footnote{AMMs have become a core market infrastructure in decentralized finance (DeFi), with major protocols such as Balancer, Curve, SushiSwap, and Uniswap deployed across major blockchains like Ethereum and Polygon. Uniswap remains a dominant protocol, representing more than 90\% of all AMM liquidity pools, with approximately 200,000 Uniswap v2 pools and 13,000 Uniswap v3 pools as of mid-2023. For complete statistics, see Dune Analytics at \url{https://dune.com/}.}
Our paper is closely related to the emerging literature on forecasting fee returns for liquidity providers.
\cite{eche:gobe:maur:26} derive an asymptotic fee formula for Uniswap v3 under specific assumptions, most notably that the prices follow the arbitrage model of \cite{angeris2021analysis}. 
\cite{agarwal2025optimal} leverage this formula to obtain the maximal fee return rate when the LP optimally exits the pool, in the sense of optimal stopping, within a geometric Brownian motion framework. 
To the best of our knowledge, no other attempt has been made to calibrate the term structure of fee rates to option prices.

Our paper also relates to existing work that characterizes the nonlinear payoff structure of liquidity positions and the exposure of LPs to impermanent loss and rebalancing risk \citep{angeris2021analysis,Clark2021,Evans2020,loesch_impermanent_2021,lvr,cartea_liq_2022,deng2023static,fukasawa2023weighted,lehar2023fragmentation,fukasawa2024modelfree,hasbrouck2024economic,agarwal2025optimal, lehar2025decentralized,lipton2025unified,eche:gobe:maur:26}. 
Our contribution is distinct. Rather than focusing on realized outcomes alone, we recover the forward-looking compensation for liquidity provision embedded in derivative prices. 
This delivers a new object---the term structure of implied fee rates---that can be used to evaluate liquidity positions, compare range designs, study the time variation in LP compensation, and connect AMM economics to broader questions in financial markets. 
 
Lastly, our paper relates to the derivatives literature by using option-implied information to construct a forward-looking measure of expected LP compensation. 
In the empirical implementation, we focus on the ETH--USDC market and calibrate a stochastic-volatility jump-diffusion (SVJD) model to Deribit option prices. 
We then use the resulting risk-neutral dynamics to price the terminal value of liquidity positions in Uniswap-style pools.
In this respect, the paper builds on option-pricing methods developed in \cite{h93,Bates1996,lewis2001simple,cont2004financial}.
Our study thus bridges the gap between derivatives markets and decentralized liquidity provision.

For brevity, all proofs are relegated to the Online Appendix.

\section{Model}\label{sec:model}

\subsection{Assumptions}
Consider a liquidity provider who chooses to supply liquidity to an AMM pool on a pair of assets $(X,Y)$, where token $Y$ is assumed to be the {\it numeraire}.
The market price of token $X$ in terms of token $Y$ at time $t$, as observed on major {\it centralized crypto exchanges}, is denoted by $S_t$ and is commonly referred to as the CEX price.


\begin{description}
\item[Assumption \namedlabel{hyp:1}{{1}}:]
To simplify the exposition, we make the usual assumption in the literature that the reference price practiced inside the pool, $P_t$, also known as the DEX price, is the same as the CEX price, $S_t$.\footnote{In the Online Appendix, we compare the CEX price $S_t$ from Binance with the DEX price $P_t$ from Uniswap v3 for the trading pair (USDC, ETH). As shown in Figure~\ref{fig:ptst}, the two prices are highly correlated, supporting the plausibility of Assumption  \ref{hyp:1}. This assumption is also consistent with the arbitrage model of \cite{angeris2021analysis}.}
\end{description}

For simplicity, we assume that the instantaneous short-term interest rates $r^X$ and $r^Y$ associated with the two tokens are constants.\footnote{These rates can be obtained from Aave, the biggest lending-borrowing protocol  \citep{gobet:hal-05544344}. Figure \ref{fig:rate:Aave} in the Online Appendix depicts the evolution of lending and borrowing rates for a few tokens. 
}


\begin{description}
\item[Assumption \namedlabel{hyp:2}{{2}}:]
We assume the existence of a unique risk-neutral measure $\bQ^Y$ associated to the numeraire $Y$,  calibrated to, and thus implied by, liquid call/put options traded in the CEX market, under which other payoffs can be consistently priced.
 \end{description}
It follows that $(e^{-(r^Y-r^X)t}S_t)_{t\ge 0}$ is a martingale under the risk-neutral measure $\bQ^Y$.
Throughout the remainder of the paper, we impose Assumptions \ref{hyp:1} and \ref{hyp:2}. 
As usual, we denote by $\bE_t^{\bQ^Y}[.]$ the conditional expectation $\bQ^Y$ given the information available at time $t$, modeled by a right-continuous complete filtration $(\mathcal{F}_t)_{t\ge 0}$.

At the entry time $t$ when the LP mints an amount of liquidity $\Delta L_\cR$ in the pool in the range $\cR=[p_a,p_b)$, accounting both for Uniswap v2 ($0=p_a<p_b=+\infty$) and Uniswap v3 ($0<p_a<p_b<+\infty$), the value of a liquidity provision $\Vlp_{t}$ is simply the pool's cash value, which is a concave positive non-decreasing function $g_\cR$ of the DEX price $P_t$ and determined by the (exogenous) AMM's reserve curve:
\begin{align*}
\Vlp_t = \Delta L_\cR\cdot g_{\cR}(P_t).
\end{align*}

We provide the explicit function $g_\cR$ for Uniswap v2 and v3 in Section \ref{sec:uniswapV3}. 
In addition, these equations are valid even if other LPs have minted/burnt liquidity in $[t,T]$ and swap traders have executed trades: this is rigorously established in \cite{eche:gobe:maur:26} for Uniswap v3, while for Uniswap v2, it is a realistic approximation.\footnote{For the protocol Uniswap v2, the fees are added to the pool, and they slightly modify the LP liquidity; however, the modification is small since the fee rate is small.} 
In addition, it suffices to consider the case of unit liquidity $\Delta L_\cR=1$, since all values are multiples of $\Delta L_\cR$, as it is explained in detail in \cite{eche:gobe:maur:26}.
Hence, hereafter, we set 
\begin{align}\Delta L_\cR=1.
\label{eq:unit:liquidity}
\end{align}
However, for a given investment horizon $T > t$, the value of her liquidity position $\Vlp_T$ is not simply the pool cash value $g_\cR(P_T)$ but rather the sum of the pool cash value $g_\cR(P_T)$ and the $Y$-value $\Gamma^{\cR}_{t,T}$ of fees  in tokens $X$ and $Y$ collected over the interval $[t,T]$, i.e.
\[\Vlp_T = g_\cR(P_T) + \Gamma^{\cR}_{t,T}.\footnote{Naturally, since $\lim_{T\rightarrow t}\Gamma^{\cR}_{t,T} = \Gamma^{\cR}_{t,t} = 0$, it follows directly that 
\[\lim_{T\rightarrow t}\Vlp_T =\Vlp_t. \]}\]

As previously discussed, our main objective is to compute the {\it implied total fee return} of one unitary liquidity in the range $\cR$, defined as
\begin{align}
R_{t,T}^{\cR}=\frac{\bE_t^{\bQ^Y}[e^{-r^Y(T-t)}\Gamma^{\cR}_{t,T}]}{g_\cR(S_t)}. 
\end{align}
We refer to this object as a total fee return because it measures the accumulated expected fees over the full holding period $[t,T]$, normalized by the initial value of the liquidity position.

From the absence of arbitrage and the risk-neutral valuation, we can relate values at times $t$ and $T$.
\begin{prop}\label{prop:accumulated_fees}
The implied accumulated fees is given by
 \begin{align}\label{eq:accumulated_fees}
\bE_t^{\bQ^Y}[e^{-r^Y(T-t)}\Gamma^{\cR}_{t,T}]= g_\cR(S_t)  - \bE_t^{\bQ^Y}[e^{-r^Y(T-t)}g_\cR(S_T)],
\end{align}
which implies the following formula for the implied total fee return:
\begin{align}
R_{t,T}^{\cR} = 1  - \frac{\bE_t^{\bQ^Y}[e^{-r^Y(T-t)}g_\cR(S_T)]}{g_\cR(S_t)}.\label{eq:gamma} 
\end{align}
\end{prop}


We are then able to analyze the bounds for $R_{t,T}^{\cR}$. The next proposition presents the result. 

\begin{prop}\label{prop:bounds}
Under the assumptions of non-negative interest rates $r^X, r^Y\ge 0$, the average accumulated fees are always nonnegative, that is,  
\begin{align}\label{eq:accumulated_fees:2}
 0\leq  \bE_t^{\bQ^Y}[e^{-r^Y(T-t)}\Gamma^{\cR}_{t,T}] < g_\cR(S_t),
 \end{align}
 implying $R_{t,T}^{\cR} \in [0,1)$
\end{prop}

Proposition \ref{prop:accumulated_fees} shows that computing the implied total fee return $R_{t,T}^{\cR}$ reduces to determining the value of
\[
H_{t,T}\equiv \bE_t^{\bQ^Y}[e^{-r^Y(T-t)}g_\cR(S_T)].
\]
The key insight is that $H_{t,T}$ is simply the time-$t$ price of a claim with terminal payoff $g_\cR(S_T)$. 
Under this interpretation, the claim can be valued using liquid vanilla calls and puts written on the CEX price $S$ under a calibrated model. Once $H_{t,T}$ is known, the implied total fee return follows directly from Eq.\eqref{eq:gamma}. 

\subsection{Bonding curve $g_\cR$ on Uniswap}\label{sec:uniswapV3}
In this section, we derive analytical expressions for the implied total fee return using the relationship in Eq.\eqref{eq:gamma}. 
The expression for $\Vlp_t=g_\cR(P_t)$ in terms of the DEX price $P_t$ depends on the AMM design. We start by analysing Uniswap v3 with a finite range $\cR=[p_a,p_b)$, since Uniswap v2 follows by taking the limit $p_a\downarrow 0$ and $p_b\uparrow +\infty$.

The AMM Uniswap v3 is summarized in \cite{Adams2021}, with detailed mechanisms described in \cite{eche:gobe:maur:26} starting from the open source code.\footnote{See \url{https://github.com/Uniswap/v3-core/}.}
In Uniswap v3 pools, the pool price can take on any value, but liquidity is concentrated on price ranges denoted by $\cR=[p_a,p_b)$.
Fees owed to each LP are tied to the price range over which the LP minted liquidity and accrue only when the pool price evolves within that range.
The bounds of these ranges $[p_a,p_b)$ are a discrete set of price values that are parametrized in the form $p(\Delta) = 1.0001^{\Delta}$, where the tick index $\Delta$ is a multiple of a fixed tick spacing $\Delta_\text{tick}$. 
The tick spacing $\Delta_\text{tick}$ is determined at the setting of the pool and depends on the swap fee tier $\phi$.
In the default setting, the tick spacing is $\Delta_\text{tick}=10,60,200$ for swap fee tiers $\phi=0.05\%,0.3\%,1\%$, respectively, although the smart contract also allows pools to be created with other tick spacings and fee tiers.


Under Uniswap v3, and using the convention that $P_t$ denotes the price of token $X$ in units of token $Y$, the bonding curve satisfied by the reserve quantities $(x_t,y_t)_{t\geq0}$ of the asset pair $(X,Y)$ is
\begin{equation} \label{eq:v3}
\left(x_t + \frac{L_\cR}{\sqrt{p_b}}\right)
\left(y_t + L_\cR\sqrt{p_a}\right) = L_\cR^2,
\end{equation}
where $L_\cR>0$ denotes the amount of liquidity available in the price range $\cR$ at the time considered.

As described in \cite{Adams2021} and \cite{eche:gobe:maur:26}, the reserve holdings \( (x_t, y_t) \) in the price range $\cR$ can be expressed as functions of the reference price $P_t$ as follows:
\begin{equation} \label{eq:holdings}
\begin{aligned}
(x_t,y_t) =&
\begin{cases}
\left(L_\cR \left( \displaystyle\frac{1}{\sqrt{p_a}} - \displaystyle\frac{1}{\sqrt{p_b}} \right), 0\right), & \text{if } P_t < p_a, \\[10pt]
\left(L_\cR \left(\displaystyle\frac{1}{\sqrt{P_t}} - \frac{1}{\sqrt{p_b}} \right), L_\cR \left( \sqrt{P_t} - \sqrt{p_a} \right)\right), & \text{if } p_a \leq P_t \leq p_b, \\[10pt]
\left(0,L_\cR \left( \sqrt{p_b} - \sqrt{p_a} \right)\right) & \text{if } P_t > p_b,
\end{cases}\\
=&\left(L_\cR \left(\displaystyle\frac{1}{\sqrt{p_a\lor P_t\land p_b}} - \frac{1}{\sqrt{p_b}} \right), L_\cR \left( \sqrt{p_a\lor P_t\land p_b} - \sqrt{p_a} \right)\right).
\end{aligned}
\end{equation}
If an LP supplies an amount of liquidity $\Delta L_\cR$ over this range, then, regardless of subsequent liquidity provision by other LPs or trades by swap traders, the LP owns the following quantities of tokens \cite[Proposition 2.2]{eche:gobe:maur:26}:
\begin{equation} \label{eq:holdings:LP}
(\Delta x_t,\Delta y_t) =\left(\Delta L_\cR \left(\frac{1}{\sqrt{p_a\lor P_t\land p_b}} - \frac{1}{\sqrt{p_b}} \right), \Delta  L_\cR\left( \sqrt{p_a\lor P_t\land p_b} - \sqrt{p_a} \right)\right).
\end{equation}
Assuming now a unit of liquidity as in Eq.\eqref{eq:unit:liquidity},
the value of the LP position denominated in token $Y$ is then
\begin{align} 
\Vlp_{t} &= \Delta x_t \cdot P_t + \Delta y_t \\
&=\frac{P_t}{\sqrt{p_a\lor P_t\land p_b}} - \frac{P_t}{\sqrt{p_b}} +\sqrt{p_a\lor P_t\land p_b} - \sqrt{p_a}\\
&=:g_\cR^{\text{Uniswap v3}}(P_t).\label{eq:vt:0:uniswapv3}
\end{align}
An interesting special case arises when $p_a \downarrow 0$ and $p_b \uparrow +\infty$, in which case the Uniswap v3 payoff converges to the Uniswap v2 expression\footnote{This limit is valid for $g_\cR^{\text{Uniswap v2}}(\cdot)$ itself. The corresponding pricing formula requires separate treatment; see the remark in the proof of Proposition \ref{prop:bates} in the Online Appendix.}
\begin{align} 
\Vlp_{t} = 2\sqrt{P_t}
=:g_\cR^{\text{Uniswap v2}}(P_t).\label{eq:vt:0:uniswapv2}
\end{align}

These expressions confirm that $g_\cR$ is concave, positive, and non-decreasing. 
This concavity has a well-known consequence: compared to a buy-and-hold strategy, commonly referred to as a HODL strategy, the liquidity providing strategy incurs impermanent losses.
Suppose an investor adopts a HODL strategy at time $t$ by holding the same quantities $(\Delta x_t,\Delta y_t)$ that the LP would otherwise have deposited, as given in Eq.\eqref{eq:holdings:LP}, so that $\text{HOLD}_t = \Delta x_t\cdot S_t + \Delta y_t = g_\cR(S_t)$.
At time $T$, the value of the HODL strategy becomes $\text{HOLD}_T = \Delta x_t\cdot S_T + \Delta y_t$, the tangent line to $g_\cR$ at $S_t$.
The difference between the values of the pool position $g_\cR$ (without fees) and the HODL portfolio is known in the literature as {\it impermanent loss}:
 \[\text{IL}_{t,T} =  g_\cR(S_T) - \text{HOLD}_T.\]
Since $g_\cR$ is concave, it lies weakly below its tangent line at every point, so, in the absence of fees, impermanent losses are always non-positive, that is, $\text{IL}_{t,T}\leq 0$.

In contrast to HODL strategies, our replicating strategy using vanilla options fully hedges the concave component of the pool cash value and generates a non-negative discounted fee revenue, quantifying the implied fees (i.e., the expected risk-neutral fees) to the market participants so they have more information when deciding to provide liquidity.

\begin{figure}[h]
    \centering
    \includegraphics[width=0.7\textwidth]{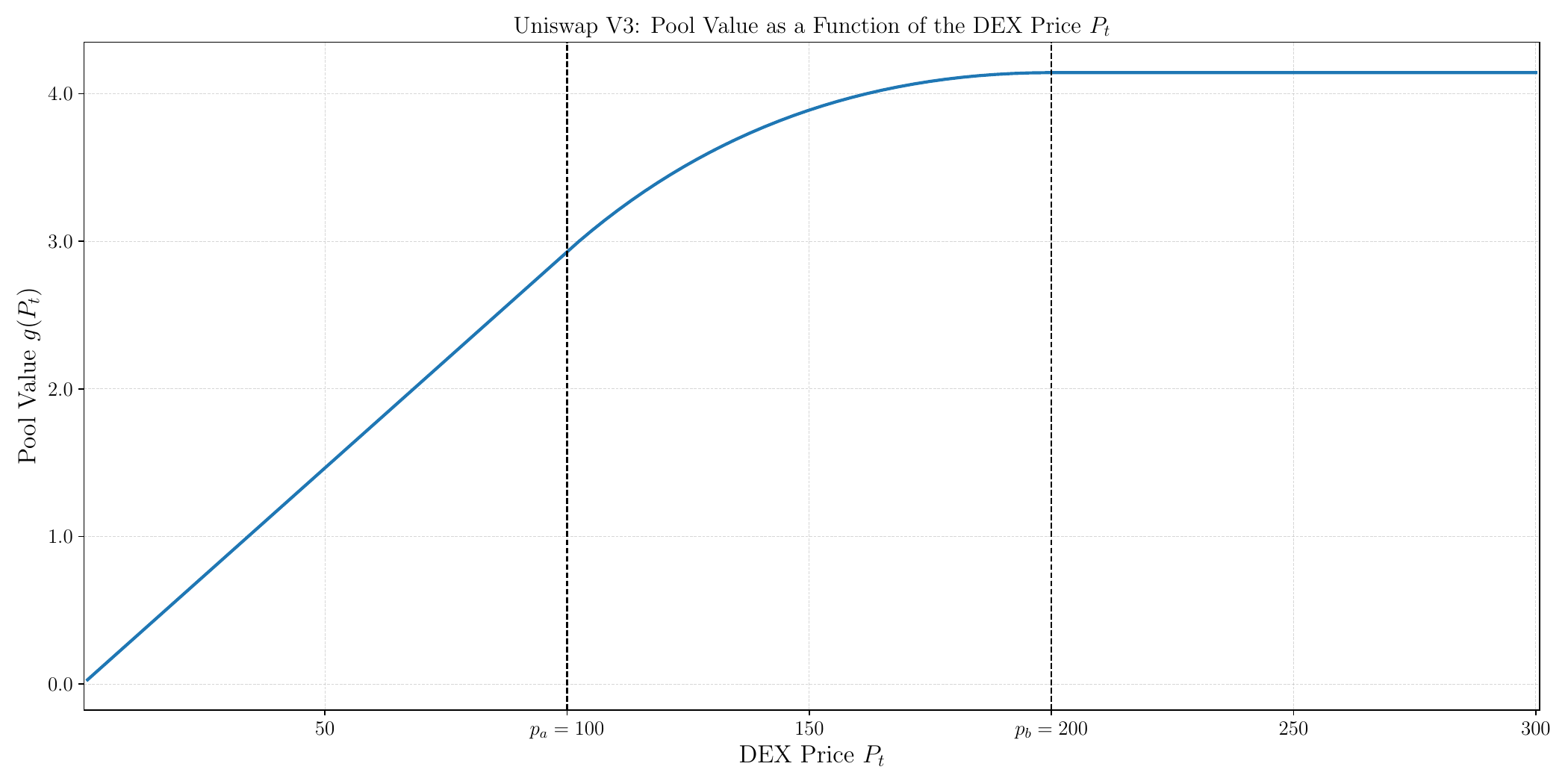}    
    \caption{\footnotesize The graph shows the pool value $\Vlp_t=g_\cR(P_t)$ in the framework of Uniswap v3, described in Eq.\eqref{eq:vt:0:uniswapv3}, as a function of the DEX price $P_t$. We set $(p_a,p_b) = (100,200)$ and vary the DEX price $P_t$ in the interval $[1, 300]$. 
    As discussed in \cite{lamb:perp:21}, the payoff is concave over the active liquidity range and capped above $p_b$, resembling a covered-call payoff: the LP remains exposed to downside movements below the range but gives up upside exposure beyond the upper price bound in exchange for fee compensation. See the discussion in \citet[Section 3.3]{eche:gobe:maur:26}.}
    \label{fig:uniswapv3}
\end{figure}

Figure \ref{fig:uniswapv3} illustrates the shape of the pool value function in Eq.\eqref{eq:vt:0:uniswapv3} as a function of the DEX price $P_t$.
Under Assumption \ref{hyp:1}, the market value of the portfolio $\Vlp_t$ is fully characterized as a function of the CEX price $S_t$. We then proceed to compute the remaining component of Eq.\eqref{eq:gamma}, that is,  $H_{t,T} = \bE^{\bQ^Y}_t\left[e^{-r^Y(T-t)}g_\cR(S_T)\right]$.

\subsection{Explicit computations in the SVJD model}
\label{subsection:cond:exp:SVJD}
To compute $\bE_t^{\bQ^Y}[e^{-r^Y(T-t)}g_\cR(S_T)]$, we assume that the underlying asset price follows an SVJD model, as in \citet{Bates1996}. We adopt this specification because it is flexible enough to capture salient features of cryptocurrency prices while still delivering semi-closed-form valuation formulas. 
It is important to emphasize, however, that our framework is not limited to the SVJD model. More generally, any option-pricing model capable of valuing the claim $g_\cR(S_T)$ can be used instead.\footnote{According to the replication argument of \cite{carr_madan_static}, a terminal payoff $g_\cR(S_T)$ can, in principle, be replicated by an infinite number of vanilla calls and puts. Thus, if option prices were available for a continuum of strikes, one could recover $H_{t,T}=\bE_t^{\bQ^Y}[e^{-r^Y(T-t)}g_\cR(S_T)]$ directly from market prices without specifying a parametric model. 
In practice, however, option prices are observed only on a finite and irregular grid of strikes and maturities. 
Since our objective is to construct implied total fee returns across different horizons and liquidity ranges, we use a calibrated option-pricing model to obtain a smooth risk-neutral distribution over the relevant set of maturities and terminal prices. 
The SVJD specification provides a flexible and tractable way to perform this interpolation and extrapolation while capturing stochastic volatility and jump risk, which are salient features of cryptocurrency option markets.}

\cite{Bates1996} assumes that under the risk-neutral probability 
$\bQ^Y$, the underlying asset price $S_t$ and the stochastic volatility $V_t$ follow the SDEs:
\begin{align}\label{eq:svjd}
\begin{aligned}
\dS_t/S_t &= (r^Y-r^X)\dt-\lambda\overline{k}\dt + \sqrt{V_t}\dZ_t + k \dq_t,\\
\dV_t &= (b_v - \beta V_t)\dt + \sigma_v\sqrt{V_t}\dZ^v_t,\\
{\rm cor}(\dZ_t,\dZ^v_t)&=\rho \dt,\quad {\rm prob}(\dq_t=1)=\lambda \dt,\\
\ln(1+k)&\sim N\left(\kappa-\frac{1}{2}\delta^2,\delta^2\right).
\end{aligned}
\end{align}
Remarkably, the characteristic function of the log-price
$$G(z, S, V, \theta)=\bE^{\bQ^Y}\left[e^{z \log(S_T)}\mid S_t=S,V_t=V\right],\qquad z\in \bC, \quad \theta=T-t,$$
is known analytically.\footnote{See \cite[Section 4]{duffiepansingleton2000} for the explicit formula.}
It is finite when $\Re(z)\le 1$. 
Moreover, $\sup_{z=\frac 12+iu,u\in \bR} |G(z, S, V, \theta)|<+\infty$, so that $G$ is uniformly bounded along the line $\Re(z)= 1/2$.\footnote{This property is used to justify the dominated convergence argument in the proof of Proposition \ref{prop:bates}.}

Moreover, the payoff expressed in Eq.\eqref{eq:vt:0:uniswapv3} is a generalized power option with $\alpha=1/2$, which allows us to obtain a semi-analytical expression for $H_{t,T}$ using the Fourier approach of \citet{lewis2001simple}.
The next proposition presents the results.

\begin{prop}\label{prop:bates}
Consider the SVJD model as in Eq.\eqref{eq:svjd}, and $g_\cR$ the payoff function described in Eq.\eqref{eq:vt:0:uniswapv3} in Uniswap v3.
The price function $$H(\theta, S, V) := \bE^{\bQ^Y}\left[e^{-r^Y(T-t)}g_\cR(S_T)\mid S_t=S,V_t=V\right]$$ is given by
\begin{align} 
&H(\theta, S, V)=e^{-r^Y\theta} \frac{2}{\pi} \int_\bR 
e^{-i\frac{u}{2 }\log(p_ap_b)}\,
\frac{\sin\left(\frac{u}{2 }\log(p_b/p_a)\right)}{u(1+4u^2)}
G(1/2+iu, S,V,\theta)\dd u.
\end{align}
Moreover, in Uniswap v2 ($p_a\downarrow 0$, $p_b\uparrow +\infty$), the price function simplifies to
\begin{align}\label{eq:H:v2}
H(\theta,S,V) = 2\,e^{-r^Y\theta}\,G\!\left(1/2,S,V,\theta\right).
\end{align}
\end{prop}

\section{Results}\label{sec:results}

We use the result in Proposition \ref{prop:bates} to calibrate implied fee rates in liquidity pools using option market data.
Specifically, we collect ETH-USDC option prices from Deribit on multiple dates and use them to estimate the model.
For each calibration date, we use daily option price quotes observed over the UTC trading day.
Our sample of ETH-USDC options spans the period from April 1, 2019, to May 17, 2023.
The calibration consists of obtaining the parameter vector $\Theta$, which governs the joint dynamics of the underlying asset price and its stochastic volatility process described in Eq.\eqref{eq:svjd}, by minimizing the option pricing error; for simplicity, we set $r^X=r^Y=0$.\footnote{We estimate $\Theta$ by minimizing the squared distance between model-implied prices, $H^{\text{model}}(\Theta)$, and observed market prices, $H^{\text{market}}$.
The calibration objective is therefore given by
\begin{equation}
    \min_{\Theta} \sum_{i=1}^{n} \left( H_i^{\text{market}} - H_i^{\text{model}}(\Theta) \right)^2,
\end{equation}
where $n$ denotes the number of options used in the calibration. We use the optimization algorithm of \cite{nelder_simplex_1965} to solve the minimization problem.
}  
For instance, the calibrated SVJD model parameters using options traded on April 1, 2023, with maturities ranging from 13 to 320 days, are
\begin{align}
\Theta :=& \left[\lambda, \kappa, b_v, \beta, \sigma_v, V_0, \rho, \delta, r^X, r^Y\right]\\
=&\left[0.004, 1.96,21.335, 55.3, 12.664, 0.497,  -0.124, 0.139, 0, 0 \right].
\end{align}

\paragraph{Implied fee return term structure.}
For a given tenor $ \theta=T-t$ and the implied total fee return $R^{\cR}_{t,t+\theta}$ described in Proposition \ref{prop:accumulated_fees}, define the associated continuously compounded fee rate as
\begin{align}\label{eq:y}
y(t,\theta):=\frac{1}{\theta}\log\left(1 + R^{\cR}_{t,t+\theta}\right).
\end{align}
Throughout our sample, $R^{\cR}_{t,t+\theta}>0$ for all calibration dates and liquidity ranges considered, so $y(t,\theta)$ is well defined everywhere it is reported below.

\begin{figure}[ht!]
    \centering
    \includegraphics[width=\textwidth]{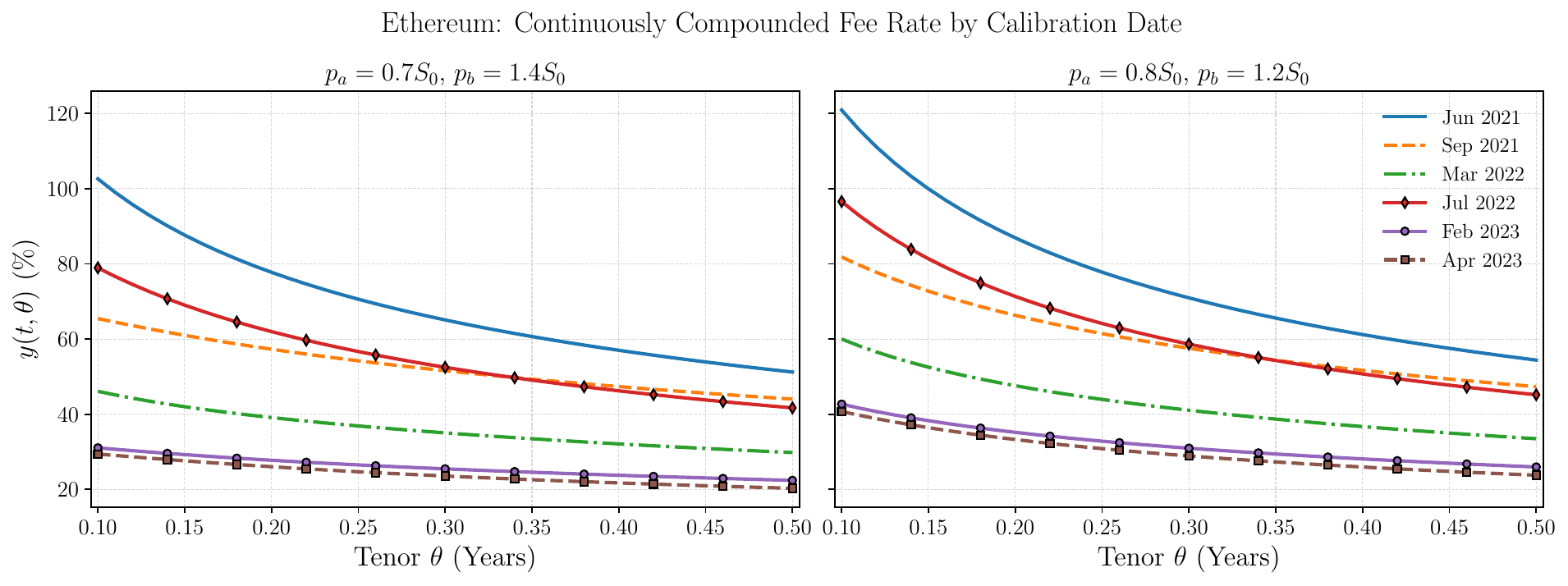}
    \includegraphics[width=\textwidth]{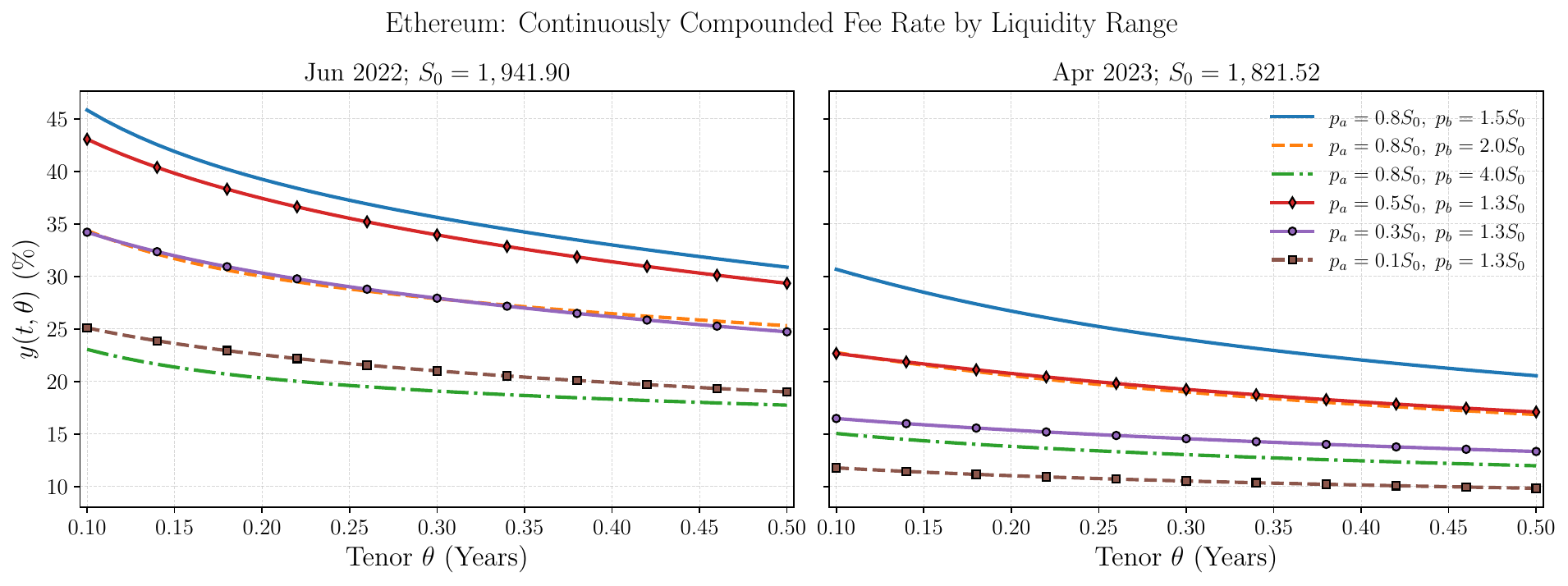}
    \caption{\footnotesize
    The top two panels plot the term structure of the continuously compounded fee rate $y(t,\theta)$ for the ETH-USDC pair across several calibration dates, with maturities ranging from 0.10 to 0.50 years. The top-left panel considers a liquidity range, $p_a=0.7S_0$ and $p_b=1.4S_0$, while the top-right panel considers a tighter range, $p_a=0.8S_0$ and $p_b=1.2S_0$. 
    The bottom two panels plot the term structure of the continuously compounded fee rate $y(t,\theta)$ for the ETH-USDC pair for two fixed calibration dates. The bottom-left panel reports the calibration for June 1, 2022, and the bottom-right panel reports the calibration for April 1, 2023. In each bottom panel, the six curves correspond to different liquidity ranges $(p_a,p_b)$, expressed as multiples of the initial spot price $S_0$.
    }
\label{fig:impliedfee}
\end{figure}

The plots in Figure \ref{fig:impliedfee} show that the continuously compounded fee rates $y(t,\theta)$ are shaped by both horizon and range geometry.
Across calibration dates, fee returns are generally downward sloping in maturity, but the level of the curve varies substantially over time, indicating that option-implied market conditions materially affect the compensation required to provide liquidity.

Range design also plays a central role, as illustrated by the bottom two panels of Figure \ref{fig:impliedfee}.
Tighter ranges centered around the current spot price produce systematically higher fee returns than wider or more asymmetric ranges. The economic intuition is straightforward: concentrated liquidity is more exposed to local price fluctuations and inventory rebalancing, so it requires greater risk-neutral compensation. Thus, both the market environment and the choice of $\cR$ jointly determine the term structure of implied fee rates.

\begin{figure}
\centering
\includegraphics[width=0.9\textwidth]{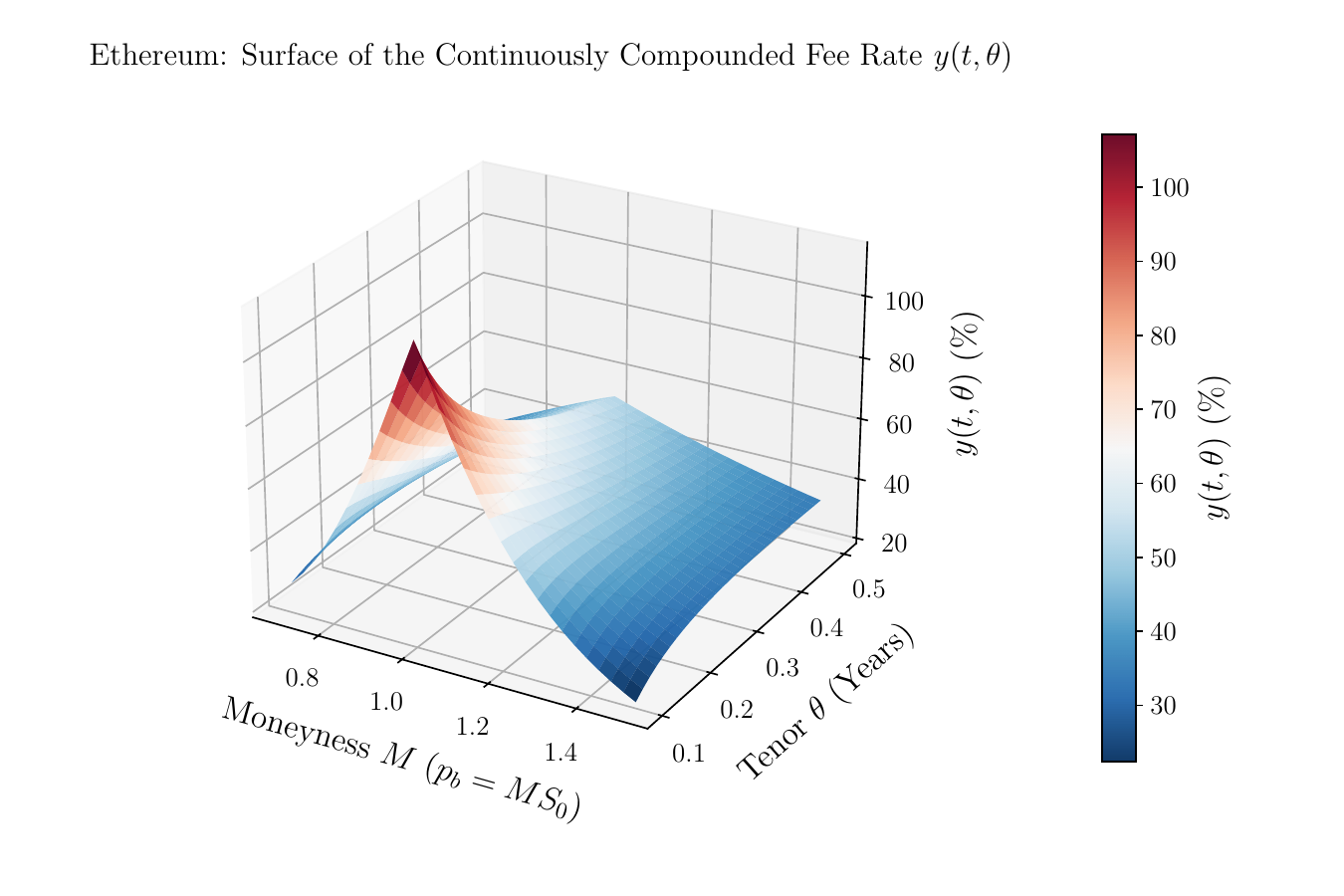}
\caption{\small Implied fee-rate surface for the ETH-USDC pair on March 1, 2022, plotted as a function of moneyness $M=p_b/S_0$ and time to maturity $T$.}
\label{fig:surface}
\end{figure}

Figure \ref{fig:surface} presents the fee-rate surface as a joint function of moneyness (as defined by $M = p_b/S_0$) and time to maturity, for a unitary range $\cR=[p_a,p_b)$, that is, the smallest possible liquidity interval in this particular pool, spanning two consecutive usable ticks.
For ranges whose upper boundary is located near the current spot price, the implied fee rate declines substantially with maturity. 
At more extreme moneyness levels, however, the term structure becomes flatter and can be upward sloping over part of the maturity range. 
The interaction between maturity and moneyness is therefore nonlinear: the effect of maturity on the implied fee rate depends on the location of the liquidity range. 
Fee rates are highest for ranges when the upper boundary of the range is near spot, particularly at short horizons, and substantially lower when it is farther from spot. 
These patterns indicate that the risk-neutral compensation for supplying liquidity depends jointly on the range geometry and the investment horizon. 

\subsection{Implied versus historical fees in Uniswap v2}
A natural question is how our theoretical term structure compares with its realized counterpart. 
To answer this question, we construct a historical accumulated fee series as follows. 
We assume an initial capital of 100{,}000 USDC invested at time $t_0$ in a liquidity position initialized at the prevailing spot price, with liquidity parameters held fixed over each evaluation window. 
Realized fees are measured directly from on-chain data and accumulated as external cash flows, without reinvestment, so that liquidity remains constant throughout the exercise.
We then compare the realized accumulated fee return with the theoretical implied total fee return computed from Eq.\eqref{eq:gamma}.

\begin{figure}[htbp]
    \centering
    \includegraphics[width=1.0\textwidth,trim=5 5 5 5,clip]{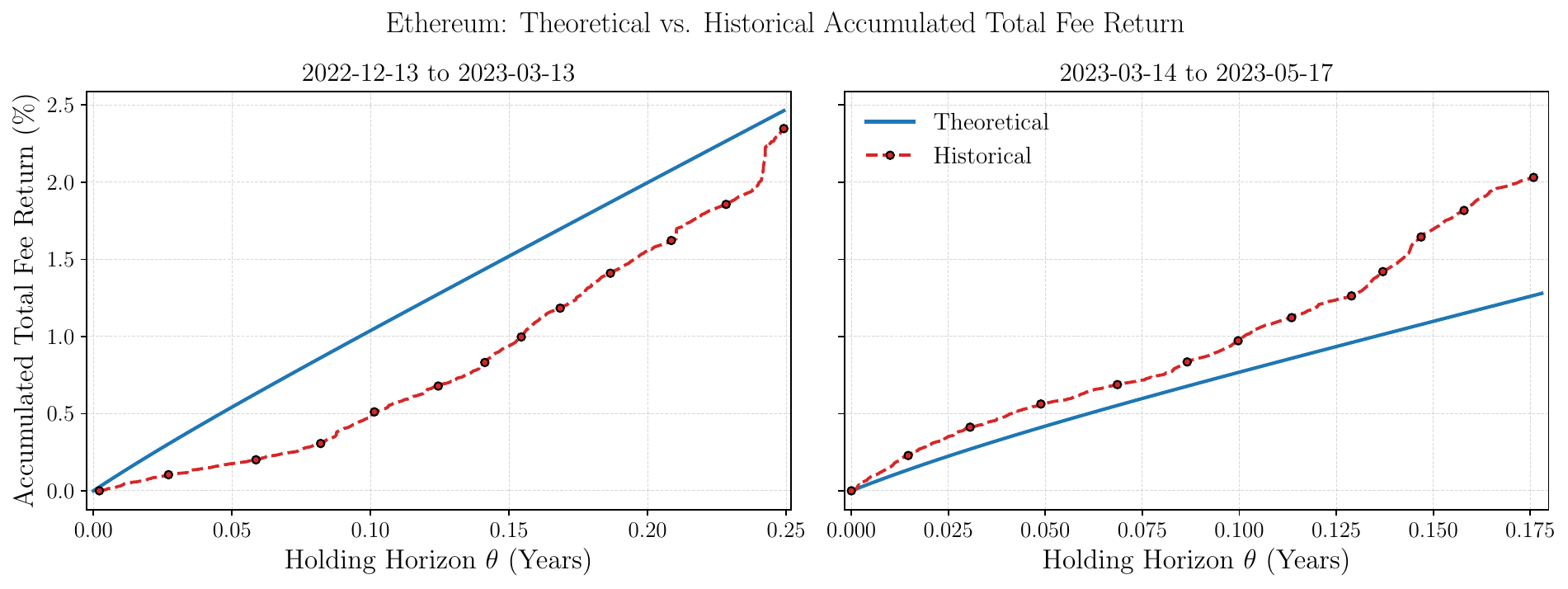}
    \caption{\footnotesize
        Theoretical and realized accumulated fees for the Uniswap~V2 ETH--USDC pool over two evaluation windows. The left panel covers December~13,~2022 to March~13,~2023, and the right panel covers March~14,~2023 to May~17,~2023. The theoretical series is based on $\bE_t^{\bQ^Y}[\Gamma^{\cR}_{t,T}] = g_\cR(S_t) - \bE_t^{\bQ^Y}[g_\cR(S_T)]$, and both series are expressed as percentages of the initial pool value.
    }
    \label{fig:lp_fees_theoretical_vs_real}
\end{figure}

Figure \ref{fig:lp_fees_theoretical_vs_real} reports the results of this exercise for two backtesting windows: December 13, 2022 to March 13, 2023 in Panel (a), and March 14, 2023 to May 17, 2023 in Panel (b).
The solid blue line represents the model-implied total fee return, while the dashed red line shows the realized historical fee accrual.
The tracking of realized fee accrual by the model-implied series suggests that the pricing-based measure captures economically meaningful variation in realized fee generation, even though the former is risk-neutral and forward-looking, whereas the latter is realized.
Deviations between the two curves are also informative, as they reflect the role of realized order flow, trading intensity, and path dependence, which need not coincide exactly with the compensation embedded in option prices at the initial date.
In sum, the close alignment between the theoretical and realized curves supports the idea that the pricing-based measure provides a useful approximation to realized fee income under a constant-liquidity strategy.

\subsection{Time series analysis of implied fee returns}

\begin{figure}[htbp]
    \centering
    \begin{subfigure}[t]{0.7\textwidth}
        \centering
        \includegraphics[width=\textwidth]{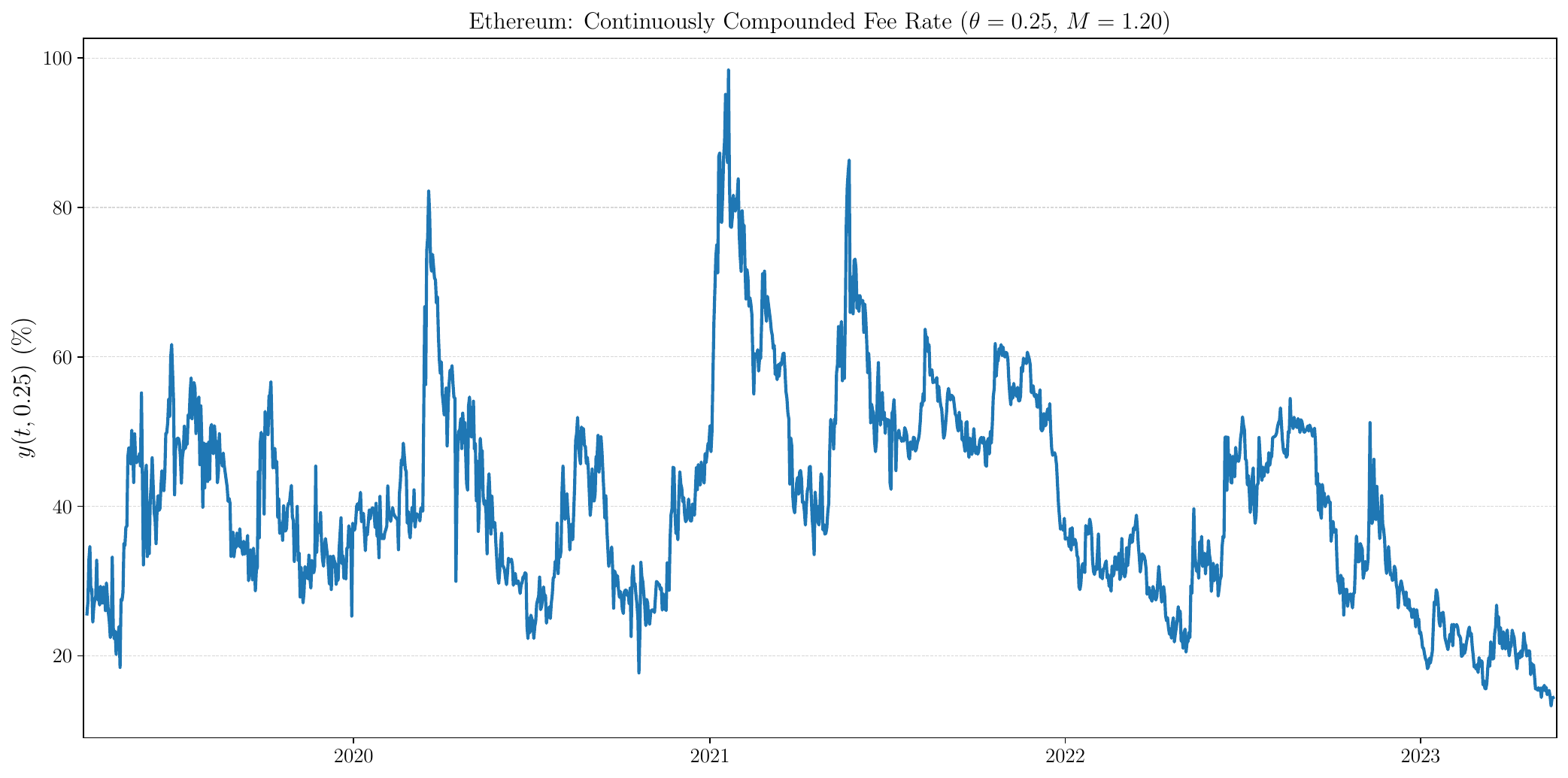}
        \caption{\footnotesize Implied time series.}
    \end{subfigure}
    \hfill\\    
    \begin{subfigure}[t]{0.7\textwidth}
        \centering
        \includegraphics[width=\textwidth]{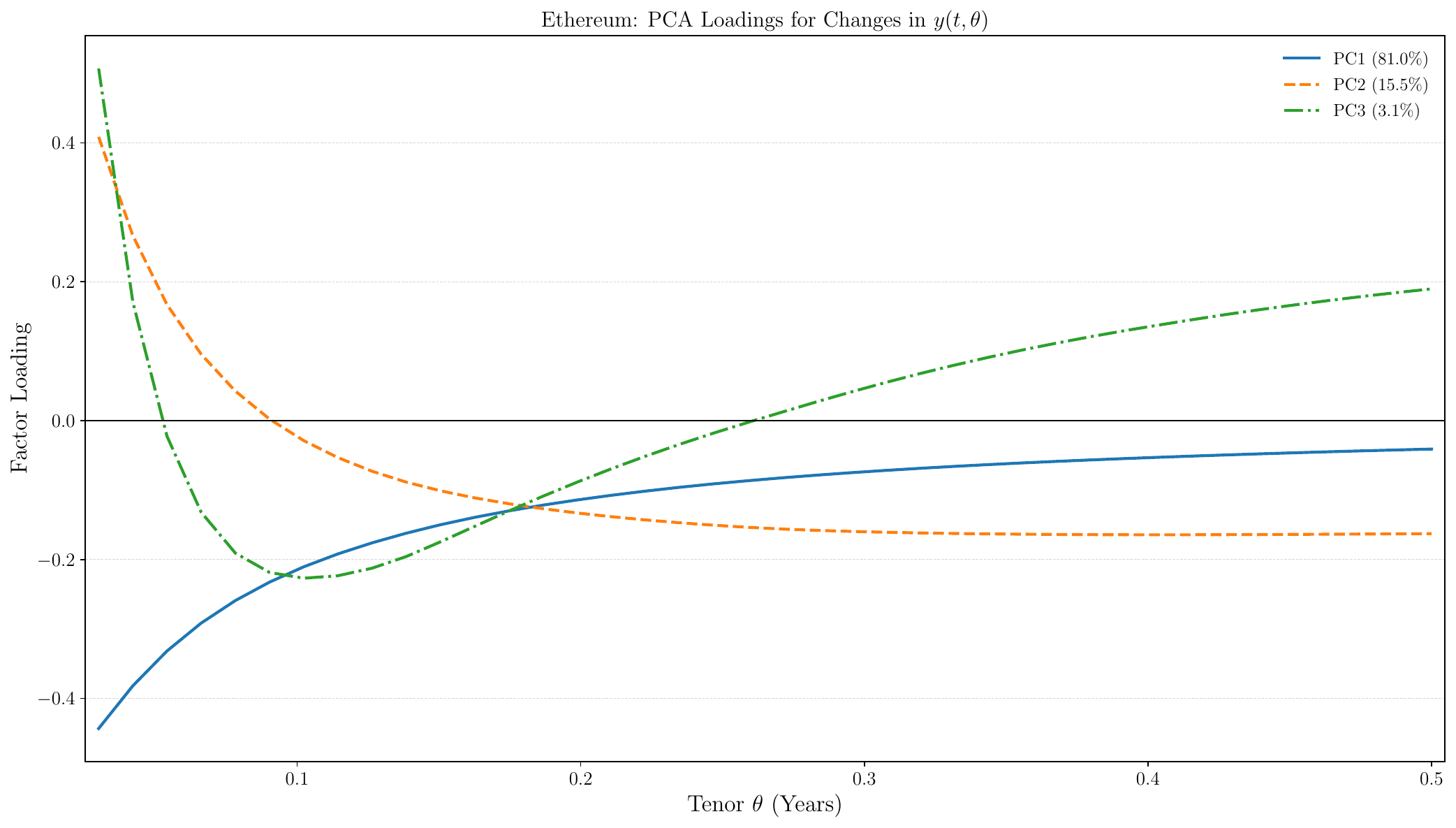}
        \caption{\footnotesize PCA loadings.}
    \end{subfigure}
    \hfill\\
    \begin{subfigure}[t]{0.7\textwidth}
        \centering
        \includegraphics[width=\textwidth]{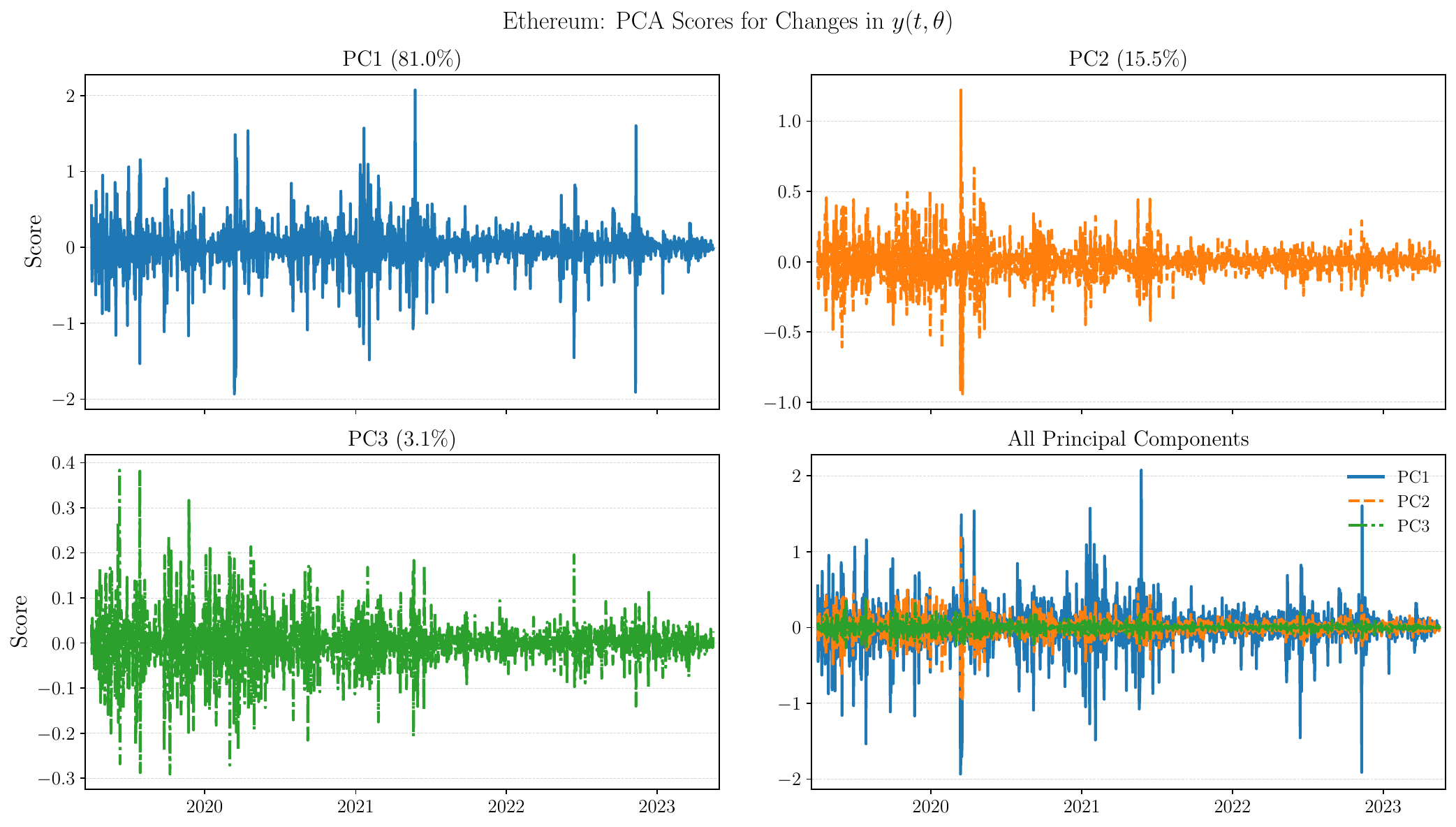}
        \caption{\footnotesize PCA factor scores.}
    \end{subfigure}
    \caption{\footnotesize Panel (a) shows the time series of implied fee returns for the ETH-USDC pair, evaluated at a fixed maturity $T=0.25$ years and a fixed liquidity-range center $M=1.20$. Panel (b) shows the principal component analysis (PCA) loadings for changes in the fee-rate curves. Panel (c) shows the PCA factor scores.}
    \label{fig:pca}
\end{figure}

Panel (a) of Figure \ref{fig:pca} reports the time series of fee rates for a fixed maturity and a fixed range center. 
The series is strongly time-varying and exhibits persistent swings, indicating that the compensation for supplying liquidity changes substantially across market states. Periods of market stress or elevated option-implied volatility tend to coincide with higher fee rates, whereas calmer periods are associated with lower values. This figure highlights an important implication of the framework: the fee rate is not a static characteristic of a pool, but rather a state-dependent quantity that moves over time with the risk-neutral distribution of future prices.

\paragraph{PCA Analysis.}
To characterize innovations in the term structure, we perform a PCA on first differences of the continuously compounded 
fee-rate curves,
\begin{align}
\Delta y(t,\theta)
=
y(t,\theta)-y(t^{-},\theta),
\end{align}
where $t^{-}$ denotes the preceding observation date. Differencing removes 
much of the persistence in fee-rate levels and isolates changes in the 
compensation implied by option prices.

Panel (b) of Figure \ref{fig:pca} shows that the first three principal components explain $99.6\%$ of the variation in 
$\Delta y(t,\theta)$, while Panel (c) shows the time series of their scores.
The first component alone accounts for $81\%$. Its 
loadings are largest at short tenors and decline monotonically with $\theta$, 
indicating that the dominant innovations are concentrated at the short end 
of the fee-rate curve. Thus, revisions in near-term compensation constitute 
the principal source of variation in implied fee rates.

The second component explains an additional $15.5\%$. Its loadings change 
sign between the shortest and longer tenors, so this factor rotates the term 
structure rather than shifting it uniformly. Positive realizations of the 
factor raise medium- and long-tenor rates relative to short-tenor rates, 
while negative realizations produce the opposite movement. We therefore 
interpret the second component as a slope factor.

The third component accounts for $3.1\%$ of the variation and exhibits 
multiple sign changes across tenors. It primarily captures curvature 
adjustments that are orthogonal to the short-end and slope movements 
represented by the first two components. Although its contribution is 
substantially smaller, it identifies localized changes in the shape of the 
fee-rate curve that cannot be summarized by a simple rotation.

The corresponding factor scores fluctuate around zero and display occasional 
large innovations. These episodes indicate that revisions to option-implied 
liquidity-provider compensation are concentrated in relatively short periods 
rather than arriving at a constant rate. Overall, the results reveal a highly 
parsimonious factor structure: two components explain $96.5\%$ of changes in 
$y(t,\theta)$, and three components explain virtually all of them. Consequently, the dynamics of changes in the fee-rate curve can be represented accurately by a small number of factors.

\section{Conclusion}
This paper develops a framework to recover implied fee rates in liquidity pools from option-implied information. The key pricing identity shows that expected accumulated fees can be written as the difference between the current pool value and the risk-neutral value of its terminal payoff. This converts the problem of measuring expected LP compensation into the pricing of a vanilla claim with payoff equal to the terminal value of the liquidity position. Using Fourier techniques, we have derived a closed-form formula for the expected fee returns, as an explicit one-dimensional integral of the characteristic function of the log-price: it allows for efficient computations.

Applying the framework to ETH-USDC options, we document three main results. First, implied fee rates are downward sloping in maturity and vary substantially over time, indicating that expected LP compensation is highly state dependent. Second, liquidity-range design matters materially: tighter ranges centered around the current spot price generate systematically higher implied fee returns than wider or more asymmetric ranges. Third, model-implied expected fees track realized fee accrual closely in backtests, suggesting that the pricing-based measure captures economically meaningful variation in realized LP profitability.

More broadly, the results show that implied fee returns constitute a forward-looking state variable for liquidity provision in automated market makers. They summarize, in a single object, how market conditions, option-implied risk, and range design jointly determine the compensation required to provide liquidity.
At the same time, the low-dimensional factor structure underlying this compensation suggests that, despite its state dependence, liquidity providers can track and hedge fee compensation without modeling the full cross-section of maturities.

\clearpage
\bibliographystyle{chicago}
\bibliography{bib_b}

@article{fukasawa2023weighted,
  title   = {Weighted Variance Swaps Hedge against Impermanent Loss},
  author  = {Fukasawa, Masaaki and Maire, Basile and Wunsch, Marcus},
  journal = {Quantitative Finance},
  volume  = {23},
  number  = {6},
  pages   = {901--911},
  year    = {2023},
  doi     = {10.1080/14697688.2023.2202708}
}

@article{fukasawa2024modelfree,
  title   = {Model-Free Hedging of Impermanent Loss in Geometric Mean Market Makers with Proportional Transaction Fees},
  author  = {Fukasawa, Masaaki and Maire, Basile and Wunsch, Marcus},
  journal = {Applied Mathematical Finance},
  volume  = {31},
  number  = {2},
  pages   = {108--129},
  year    = {2024},
  doi     = {10.1080/1350486X.2024.2404058}
}

@article{lipton2025unified,
  title   = {Unified Approach for Hedging Impermanent Loss of Liquidity Provision},
  author  = {Lipton, Alex and Lucic, Vladimir and Sepp, Artur},
  journal = {Digital Finance},
  volume  = {7},
  number  = {3},
  pages   = {429--477},
  year    = {2025},
  doi     = {10.1007/s42521-025-00144-5}
}

@article{cartea_liq_2022,
	author = {Cartea, \'{A}lvaro and Drissi, Fay\c{c}al and Monga, Marcello},
	doi = {10.1137/23M1602103},
	eprint = {https://doi.org/10.1137/23M1602103},
	journal = {SIAM Journal on Financial Mathematics},
	number = {3},
	pages = {931-959},
	title = {Decentralized Finance and Automated Market Making: Predictable Loss and Optimal Liquidity Provision},
	url = {https://doi.org/10.1137/23M1602103},
	volume = {15},
	year = {2024}}

@article{nelder_simplex_1965,
	title = {A {Simplex} {Method} for {Function} {Minimization}},
	volume = {7},
	issn = {0010-4620, 1460-2067},
	url = {https://academic.oup.com/comjnl/article-lookup/doi/10.1093/comjnl/7.4.308},
	doi = {10.1093/comjnl/7.4.308},
	language = {en},
	number = {4},
	urldate = {2024-01-07},
	journal = {The Computer Journal},
	author = {Nelder, J. A. and Mead, R.},
	month = jan,
	year = {1965},
	pages = {308--313},
}

@book{cont2004financial,
  title={Financial modelling with jump processes},
  author={Cont, Rama and Tankov, Peter},
  year={2004},
  publisher={CRC Press}
}

@misc{loesch_impermanent_2021,
	title = {Impermanent {Loss} in {Uniswap} v3},
	url = {http://arxiv.org/abs/2111.09192},
	doi = {10.48550/arXiv.2111.09192},
	urldate = {2024-07-16},
	publisher = {arXiv},
	author = {Loesch, Stefan and Hindman, Nate and Richardson, Mark B. and Welch, Nicholas},
	month = nov,
	year = {2021},
	note = {arXiv:2111.09192 [q-fin]},
}

@article{Adams2020,
   author = {Hayden Adams and Noah Zinsmeister and Dan Robinson},
   title = {Uniswap v2 Core},
   journal={Uniswap.org},
     note={\url{https://uniswap.org/whitepaper.pdf}},
   year = {2020},
}

@article{h93,
  title={A Closed-Form Solution for Options with Stochastic Volatility with Applications to Bond and Currency Options},
  author={Heston, Steven L},
  journal={The Review of Financial Studies},
  volume={6},
  number={2},
  pages={327--343},
  year={1993},
  publisher={Oxford University Press}
}

@incollection{carr_madan_static,
	address = {Cambridge},
	title = {Towards a {Theory} of {Volatility} {Trading}},
	isbn = {9780511569708},
	url = {https://www.cambridge.org/core/product/identifier/CBO9780511569708A023/type/book_part},
	urldate = {2023-09-27},
	booktitle = {Option {Pricing}, {Interest} {Rates} and {Risk} {Management}},
	publisher = {Cambridge University Press},
	author = {Carr, P. and Madan, D.},
	editor = {Jouini, E. and Cvitanic, J. and Musiela, Marek},
	year = {2001},
	doi = {10.1017/CBO9780511569708.013},
	pages = {458--476},
}

@article{lvr,
	title = {Automated {Market} {Making} and {Loss}-{Versus}-{Rebalancing}},
	copyright = {arXiv.org perpetual, non-exclusive license},
	url = {https://arxiv.org/abs/2208.06046},
	doi = {10.48550/ARXIV.2208.06046},
	urldate = {2023-09-02},
	author = {Milionis, Jason and Moallemi, Ciamac C. and Roughgarden, Tim and Zhang, Anthony Lee},
	year = {2022},
}

@report{Clark2021,
   author = {Joseph Clark},
   title = {A deconstructed constant product market},
   url = {https://ssrn.com/abstract=3856446},
   year = {2021},
}

@article{Evans2020,
	author = {Evans, Alex},
	journal = {Cryptoeconomic Systems},
	editor = {Youngblom, Reuben},
	number = {2},
	year = {2021},
	month = {oct 22},
	note = {https://cryptoeconomicsystems.pubpub.org/pub/evans-g3m-returns},
	publisher = {Metagov},
	title = {Liquidity {Provider} {Returns} in {Geometric} {Mean} {Markets}},
	volume = {1},
}

@article{angeris2021analysis,
  title={An analysis of {U}niswap markets},
  author={Angeris, Guillermo and Kao, Hsien-Tang and Chiang, Rei and Noyes, Charlie and Chitra, Tarun},
  year={2021},
  volume={1},
  number={1},
  publisher={PubPub},
  journal={Cryptoeconomic Systems}
}

@article{lehar2025decentralized,
  author  = {Lehar, Alfred and Parlour, Christine A.},
  title   = {Decentralized Exchange: The {Uniswap} Automated Market Maker},
  journal = {Journal of Finance},
  year    = {2025},
  volume  = {80},
  number  = {1},
  pages   = {321--374},
  doi     = {10.1111/jofi.13405}
}

@article{lehar2023fragmentation,
  author  = {Lehar, Alfred and Parlour, Christine A. and Zoican, Marius},
  title   = {Fragmentation and Optimal Liquidity Supply on Decentralized Exchanges},
  year    = {2023},
  note    = {arXiv:2307.13772},
  doi     = {10.48550/arXiv.2307.13772}
}

@article{Neuder2021,
   author={Zhou Fan and Francisco Marmolejo-Cossio and Daniel J. Moroz and Michael Neuder and Rithvik Rao and David C. Parkes},
   month = {6},
   title={Strategic Liquidity Provision in {U}niswap v3}, 
   journal={arXiv preprint arXiv:2106.12033},
   year = {2021},
}

@article{Bates1996,
  title={Jumps and stochastic volatility: Exchange rate processes implicit in deutsche mark options},
  author={Bates, David S},
  journal={The Review of Financial Studies},
  volume={9},
  number={1},
  pages={69--107},
  year={1996},
  publisher={Oxford University Press}
}

@article{duffiepansingleton2000,
  title={Transform analysis and asset pricing for affine jump-diffusions},
  author={Duffie, Darrell and Pan, Jun and Singleton, Kenneth},
  journal={Econometrica},
  volume={68},
  number={6},
  pages={1343--1376},
  year={2000},
  publisher={Wiley Online Library}
}

@article{deng2023static,
  author  = {Deng, Jun and Zong, Hua and Wang, Yun},
  title   = {Static Replication of Impermanent Loss for Concentrated Liquidity Provision in Decentralised Markets},
  journal = {Operations Research Letters},
  year    = {2023},
  volume  = {51},
  number  = {3},
  doi     = {10.1016/j.orl.2023.03.002}
}

@article{hasbrouck2024economic,
  author  = {Hasbrouck, Joel and Rivera, Thomas J. and Saleh, Fahad},
  title   = {An Economic Model of a Decentralized Exchange with Concentrated Liquidity},
  journal = {Management Science},
  year    = {2024},
  note    = {Articles in Advance},
  doi     = {10.1287/mnsc.2024.04510}
}

@article{Adams2021,
   author = {Hayden Adams and Noah Zinsmeister and Moody Salem  and  River Keefer and Dan Robinson},
   title = {Uniswap v3 Core},
   year = {2021},
   journal={Uniswap.org},
   note={\url{https://app.uniswap.org/whitepaper-v3.pdf}}
}

@misc{Jin2022,
  doi = {10.48550/ARXIV.2209.01653},
  url = {https://arxiv.org/abs/2209.01653},
  author = {Kuan, Jin Hong},
  title = {Liquidity Provision Payoff on Automated Market Makers},
  publisher = {arXiv},
  year = {2022},
  copyright = {Creative Commons Attribution 4.0 International}
}

@article {eche:gobe:maur:26,
  AUTHOR =	 {Mnacho Echenim and Emmanuel Gobet and Anne-Claire Maurice},
  TITLE =	 {Uniswap V3: Impermanent Loss Modeling and Swap Fees Asymptotic Analysis},
  JOURNAL =	 {SIAM J.Financial Mathematics},
  DOI =		 {10.1137/23M1606149},
  Volume =	 17,
  number =	 1,
  pages={1--40},
  YEAR =	 2026
}

@article{lamb:perp:21,
  title={Uniswap v3 {LP} Tokens as Perpetual Put and Call Options},
  author={G. Lambert},
  journal={Medium},
  note={\url{https://medium.com/@lambert-guillaume/uniswap-v3-lp-tokens-as-perpetual-put-and-call-options-5b66219db827}},
month={June},
  year={2021}
}

@article{agarwal2025optimal,
  author  = {Ankush Agarwal and Emmanuel Gobet},
  title   = {Optimal Exit from {Uniswap} v3 and Best Expected Return
             for a Liquidity Provider},
  journal    = {SSRN Working Paper 5422975},
  year    = {2025},
  url     = {https://papers.ssrn.com/sol3/papers.cfm?abstract_id=5422975},
}

@article{gobet:hal-05544344,
  TITLE = {{Modeling the risks within the protocol Aave, with an application to portfolio allocation}},
  AUTHOR = {Gobet, Emmanuel and Latournerie, Louis},
  journal = {HAL preprint, \url{https://hal.science/hal-05544344}},
  YEAR = {2026},
  MONTH = Mar}

@article{lewis2001simple,
  title={A simple option formula for general jump-diffusion and other exponential {L\'e}vy processes},
  author={Lewis, A.},
  year={2001},
  note={Available at SSRN: \url{http://dx.doi.org/10.2139/ssrn.282110}}
}
\clearpage

\clearpage
\appendix
\setcounter{section}{0}
\setcounter{subsection}{0}
\setcounter{table}{0}
\setcounter{figure}{0}
\setcounter{equation}{0}

\renewcommand{\thesection}{OA\arabic{section}}
\renewcommand{\thesubsection}{OA\arabic{section}.\arabic{subsection}}
\renewcommand{\thetable}{OA\arabic{table}}
\renewcommand{\thefigure}{OA\arabic{figure}}
\renewcommand{\theequation}{OA\arabic{equation}}
\pagenumbering{arabic}
\setcounter{page}{1}

\newpage
\section{Internet Appendix}\label{sec:oa}

\subsection{Proof of Proposition \ref{prop:accumulated_fees}}
\label{subsection:proof:accumulated_fees}
Using risk-neutral valuation arguments and Assumption 1, at time $t$ the pool cash value satisfies:
\begin{align}
\Vlp_t =& \bE_t^{\bQ^Y}[e^{-r^Y(T-t)}\Vlp_T] = \bE_t^{\bQ^Y}[e^{-r^Y(T-t)}g_\cR(S_T)] + \bE_t^{\bQ^Y}[e^{-r^Y(T-t)}\Gamma^{\cR}_{t,T}].
 \end{align}
Using that $\Vlp_t = g_\cR(S_t)$ and rearranging the equation readily leads to Eq.\eqref{eq:accumulated_fees}.\\

\subsection{Proof of Proposition \ref{prop:bounds}}
\label{subsection:proof:bounds}
Assume now that $r^X\ge 0$ and $r^Y\ge 0$. The right-hand side of Eq.\eqref{eq:accumulated_fees:2} is clear since $g_\cR$ is non-negative. To prove the left-hand side, observe that, since $g_\cR$ is concave and $e^{-(r^Y-r^X)t} S_t$ is a $\bQ^Y$-martingale, Jensen's inequality implies that
\begin{align*}
&\bE^{\bQ^Y}_t[g_\cR(S_T)] \leq g_\cR(\bE^{\bQ^Y}_t[S_T]) = g_\cR(e^{(r^Y-r^X)(T-t)} S_t).
\end{align*}
It remains to show that for any $\theta=T-t\ge 0$ and $S>0$, we have
\begin{align*}
e^{-r^Y \theta}
g_\cR(e^{(r^Y-r^X)\theta} S)\le g_\cR(S).
\end{align*}
\noindent \textbf{Case 1: $r^Y\le r^X$.} Leveraging that $g_\cR$ is positive and non-decreasing, we get
\begin{align*}
e^{-r^Y \theta}
g_\cR(e^{(r^Y-r^X)\theta} S)\le 
g_\cR(e^{(r^Y-r^X)\theta} S)
\le 
g_\cR(S).
\end{align*}
\noindent \textbf{Case 2: $r^Y\ge r^X$,} i.e. $\lambda=e^{(r^Y-r^X)\theta}\ge 1$. Use the concavity of $g_\cR$ with $g_\cR(0)=0$ to write
\[
S
=
\frac1{\lambda}(\lambda S)
+
\left(1-\frac1{\lambda}\right)0\Longrightarrow 
g_\cR(S)
\ge
\frac1{\lambda} g_\cR(\lambda S).\]
It readily follows that 
\begin{align}
e^{-r^Y\theta}
g_\cR\left(e^{(r^Y-r^X)\theta}S\right)
&\le
e^{-r^Y\theta}
e^{(r^Y-r^X)\theta}g_\cR(S)
\le
g_\cR(S).
\end{align}
The proof is complete.
\qed

\subsection{Proof of Proposition \ref{prop:bates}}
\label{section:proof:prop:bates}
\noindent\textbf{The Uniswap v3 case.} Define $\theta = T-t$. 
The Fourier strategy of \cite{lewis2001simple} consists in writing 
\begin{align}
H(\theta, S, V) &= \bE^{\bQ^Y}\left[e^{-r^Y(T-t)}g_\cR(S_T)\mid S_t=S,V_t=V\right]\\
&=e^{-r^Y\theta} \int_\bR  g_\cR(e^x) p(x)\dx
=e^{-r^Y\theta} \int_\bR e^{-x/2  } g_\cR(e^x) e^{x/2  }p(x)\dx\quad \label{eq:step1:lewis}
\end{align}
where $p$ stands for the density of $\log(S_T)$ conditionally to $S_t=S, V_t=V$. Here, we have assumed that $p$ exists and is such that $e^{\cdot/2}p(\cdot)\in L^2 (\bR, \dx)$; at the end, we will get rid of this assumption by a convolution argument. 
To transform Eq.\eqref{eq:step1:lewis}, we make use of Fourier techniques.
For a function $f\in L^1(\bR,\dx)$, define its Fourier transform by
$        \widehat{f}(u)  := \int_\bR e^{iux} f(x) \dd x$ for $u\in \bR$. 
In addition, if $f$ and $g$ are $L^2(\bR,\dx)$, then $\widehat f$ and $\widehat g$ are in $L^2(\bR,\dx)$ as well, and we have the Parseval-Plancherel equality
 \begin{align}
 \int_{\bR} f(x) g(x) \dx= \frac{1}{2\pi}\int_{\bR} \overline{\widehat{f}(u)} \widehat{g}(u)\dd u,
 \end{align}
 which, applied to Eq.\eqref{eq:step1:lewis}, gives
\begin{align}
H(\theta, S, V) 
&=e^{-r^Y\theta} \frac{1}{2\pi} \int_\bR \overline{\widehat{e^{-\cdot /2  } g_\cR(e^\cdot)}}(u) \widehat{e^{\cdot/2  }p(\cdot)}(u)\dd u\label{eq:step2:lewis}.
\end{align}
Note that 
\begin{align}
\widehat{e^{\cdot/2  }p(\cdot)}(u)
=\int_\bR e^{iux} e^{x/2  }p(x)\dx=G(\frac 12+iu, S,V,\theta).
\end{align}
It remains to consider $\widehat{e^{-\cdot /2  } g_\cR(e^\cdot)}(u)$.
First, we claim that $f:x\mapsto e^{-x /2  } g_\cR(e^x)$ is in $L^2(\bR,\dx)$: indeed, setting 
$$a=\log p_a, \qquad b=\log p_b$$ and reasoning on each interval, we observe from \eqref{eq:vt:0:uniswapv3} that 
\begin{align}
f(x)
&=
\begin{cases}
\displaystyle
\left(\frac{1}{\sqrt{p_a}}-\frac{1}{\sqrt{p_b}}\right)e^{x/2},
& x<a,\\[2mm]
\displaystyle
2-\frac{e^{x/2}}{\sqrt{p_b}}-\sqrt{p_a}\,e^{-x/2},
& a\le x\le b,\\[2mm]
\displaystyle
(\sqrt{p_b}-\sqrt{p_a})e^{-x/2},
& x>b.
\end{cases}
\end{align}
Hence, $f$ decays exponentially fast as $x\to \pm\infty$, and therefore is in $L^2(\bR,\dx)$. Furthermore, 
\[
\widehat f(u)
=
\int_{-\infty}^{a} e^{iux}f(x)\,dx
+
\int_{a}^{b} e^{iux}f(x)\,dx
+
\int_{b}^{\infty} e^{iux}f(x)\,dx.
\]
For the first and third terms,
\begin{align}
I_1&=
\left(\frac{1}{\sqrt{p_a}}-\frac{1}{\sqrt{p_b}}\right)
\frac{e^{(\frac12+iu)a}}{\frac12+iu}=
\left(1-\sqrt{\frac{p_a}{p_b}}\right)
\frac{e^{iua}}{\frac12+iu},\\
I_3&=
(\sqrt{p_b}-\sqrt{p_a})
\frac{e^{(-\frac12+iu)b}}{\frac12-iu}
=\left(1-\sqrt{\frac{p_a}{p_b}}\right)
\frac{e^{iub}}{\frac12-iu}.
\end{align}
For the middle term, for $u\neq 0$,
\begin{align}
I_2
=&
2\frac{e^{iub}-e^{iua}}{iu}-
\frac{e^{iub}-\sqrt{p_a/p_b}\,e^{iua}}
{\frac12+iu}
+
\frac{\sqrt{p_a/p_b}\,e^{iub}-e^{iua}}
{\frac12-iu}.
\end{align}
Summing the three contributions and simplifying yields
\begin{align}
\widehat f(u)
&=
\frac{e^{iua}}{\frac12+iu}
+
\frac{e^{iub}}{\frac12-iu}+
2\frac{e^{iub}-e^{iua}}{iu}-
\frac{e^{iub}}
{\frac12+iu}
-
\frac{e^{iua}}
{\frac12-iu}
\\
&=\left(e^{iub}-e^{iua}\right)
\left[
\frac{2}{iu}
-\frac{1}{\frac12+iu}
+\frac{1}{\frac12-iu}
\right]\\
&
=2i\,e^{iu(a+b)/2}
\sin\left(\frac{u(b-a)}{2}\right) \times 
\frac{-2i}{u(1+4u^2)}\\
&
=4\,e^{iu(a+b)/2}\,
\frac{\sin\left(u(b-a)/2\right)}{u(1+4u^2)}.
\end{align}
The above computations are valid for $u=0$ with 
$$\widehat f(0)=2 (b-a),$$ 
which is the natural continuity extension of $\widehat f$ at $u=0$.
Plugging this into \eqref{eq:step2:lewis}, we have obtained
\begin{align}
H(\theta, S, V) 
&=e^{-r^Y\theta} \frac{2}{\pi} \int_\bR 
e^{-iu(a+b)/2}\,
\frac{\sin\left(u(b-a)/2\right)}{u(1+4u^2)}
G(\frac 12+iu, S,V,\theta)\dd u\label{eq:step3:lewis}, 
\end{align}
as announced.
It remains to prove that the above formula remains valid when the conditional (to $S_t=S, V_t=V$) distribution of $\log(S_T)$ may not have a density.
Take a small Gaussian perturbation of $\log(S_T)$, i.e. $ \log(S_T) + N_{\epsilon}$, where $N_{\epsilon} \sim \cN(0, \epsilon)$ admits the density $\varphi_{\epsilon}(y) := \frac{ e^{{-\frac{y^2}{2\epsilon}}} }{ \sqrt{2\pi\epsilon} }$ and is independent of $S_T$. We easily justify that the conditional distribution of $ \log(S_T) + N_{\epsilon}$ is absolutely continuous w.r.t. the Lebesgue measure in $\bR$, with density $p_{\epsilon}(x)= \bE\left[  \varphi_{\epsilon} (x - \log(S_T))\right]$.
Observe that: 
    \begin{align*}
        \int_{\bR}  p_{\epsilon} ^{2} (x) e^{x} \dd x 
        & \leq \int_\bR \bE\left[ \varphi_{\epsilon} ^{2} (x - \log(S_T))\mid S_t=S, V_t=V\right] e^{x} \dd x \\
        & = \bE\left[\int_\bR e^x e^{-\frac{(x - \log(S_T))^2}{\epsilon}} \frac{\dd x}{2\pi\epsilon} \mid S_t=S, V_t=V\right]\\
                & = \bE\left[\int_\bR e^{\log(S_T)+y} e^{-\frac{y^2}{\epsilon}} \frac{\dd y}{2\pi\epsilon} \mid S_t=S, V_t=V\right] \\
                &= S e^{(r^Y-r^X)\theta} \int_\bR e^{y} e^{-\frac{y^2}{\epsilon}} \frac{\dd y}{2\pi\epsilon} <+\infty
    \end{align*}
    where the first inequality holds by Jensen's inequality. This proves that 
    $e^{\cdot/2}p_\epsilon(\cdot)\in L^2 (\bR, \dx)$. Therefore, the previous formula applies: 
\begin{align}
&\bE^{\bQ^Y}\left[e^{-r^Y\theta}g_\cR(S_Te^{N_\epsilon})\mid S_t=S,V_t=V\right]\\
&=e^{-r^Y\theta} \frac{2}{\pi} \int_\bR 
e^{-iu(a+b)/2}\,
\frac{\sin\left(u(b-a)/2\right)}{u(1+4u^2)}
G(\frac 12+iu, S,V,\theta)
 \bE\left[ e^{(\frac{1}{2} + iu) N_{\epsilon}}\right] 
 \dd u.
\end{align}
Let us pass to the limit as $\epsilon\to 0$ using the Lebesgue dominated convergence theorem: the left-hand-side converges to $\bE^{\bQ^Y}\left[e^{-r^Y\theta}g_\cR(S_T)\mid S_t=S,V_t=V\right]$ thanks to the boundedness and continuity of $g_\cR$. Regarding the right-hand-side, note that 
$\bE\left[ e^{(\frac{1}{2} + iu) N_{\epsilon}}\right] \to 1$ by direct computations and the function to integrate is dominated by $C/(1+u^2)$ as $u\to\pm\infty$.

\medskip
\noindent\textbf{The Uniswap v2 case.} Recall from Section \ref{sec:uniswapV3} that $g_\cR^{\text{Uniswap v2}}(P) = 2\sqrt{P} = 2e^{\frac12\log P}$ for all $P>0$, with no truncation at finite bounds. Hence, directly from the definition of $G$,
\begin{align*}
H(\theta,S,V)
&= \bE^{\bQ^Y}_t\left[e^{-r^Y\theta}\cdot 2\sqrt{S_T}\mid S_t=S,V_t=V\right]\\
&= 2e^{-r^Y\theta}\,\bE^{\bQ^Y}_t\left[e^{\frac12\log S_T}\mid S_t=S,V_t=V\right]
= 2e^{-r^Y\theta}\,G\!\left(\tfrac12,S,V,\theta\right),
\end{align*}
which is finite since $\Re(\tfrac12)=\tfrac12\le 1$, as established in Section \ref{subsection:cond:exp:SVJD}.
Unlike the Uniswap v3 case, no Fourier inversion is required: \eqref{eq:H:v2} follows directly from the definition of $G$, since the Uniswap v2 payoff has no kinks to Fourier-transform.

\paragraph{Remark.} Formula \eqref{eq:H:v2} is not directly obtained by letting $p_a\downarrow 0,\ p_b\uparrow+\infty$ inside the Fourier-integral representation of Proposition \ref{prop:bates}, since the term $\sin(u(b-a)/2)$ oscillates increasingly fast as $b-a\to\infty$ and then this limit would have to be taken in a distributional sense; instead, \eqref{eq:H:v2} is derived directly from the definition of $G$, exploiting the fact that the Uniswap v2 payoff is smooth, with no kinks or slope discontinuities to complicate the Fourier transform.

The proof of Proposition \ref{prop:bates} is complete.
\qed

\newpage

\subsection{Additional Figures}

\begin{figure}[h]
    \centering
    \includegraphics[width=0.8\textwidth]{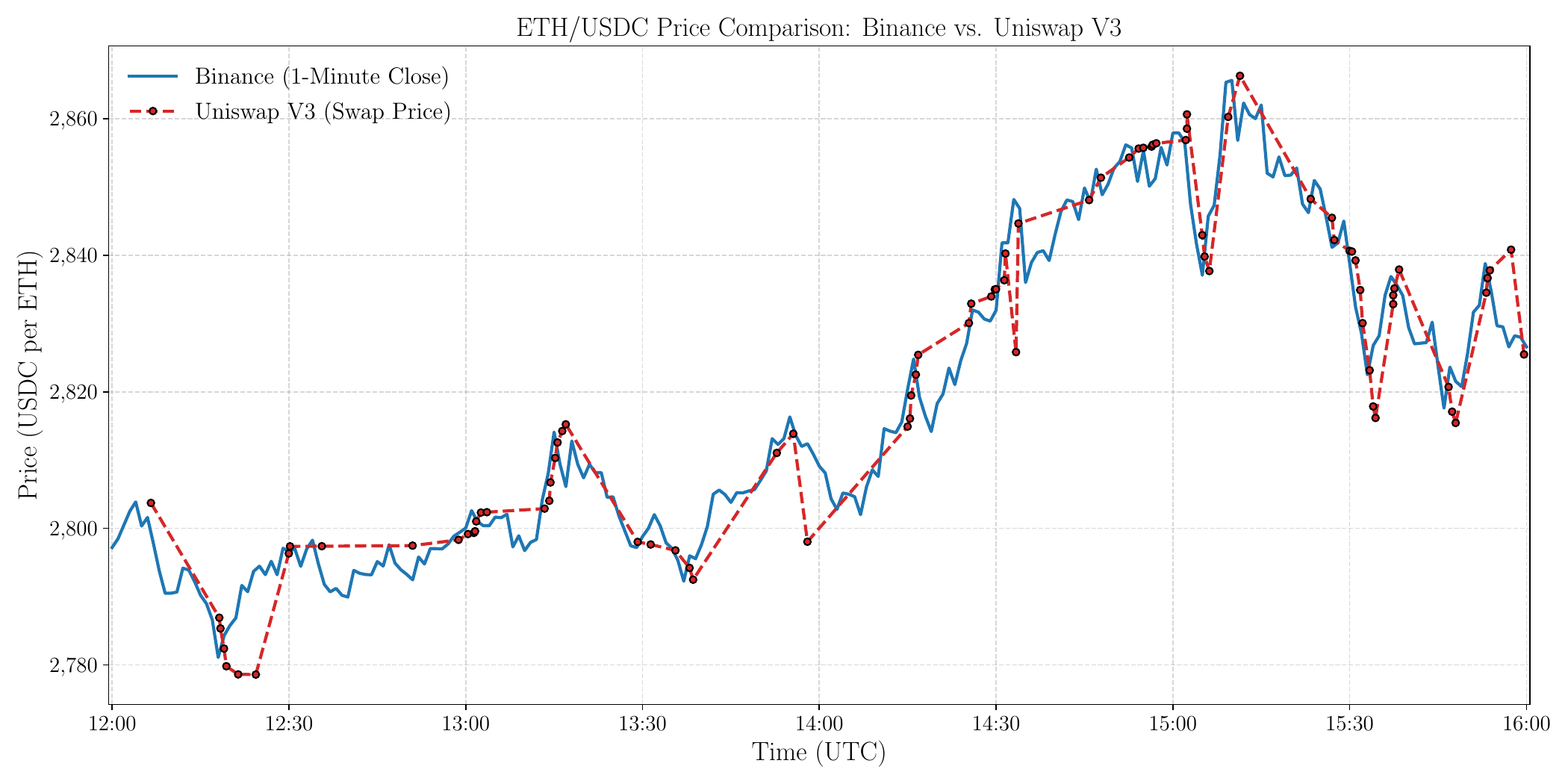}
    \caption{\small Time series of the DEX price $P_t$ from Uniswap v3 and the CEX price $S_t$ from Binance for the ETH-USDC pair on February 15, 2024, from 12:00 to 16:00 UTC.}
    \label{fig:ptst}
\end{figure}

\begin{figure}[h]
    \centering
\includegraphics[width=0.49\linewidth]{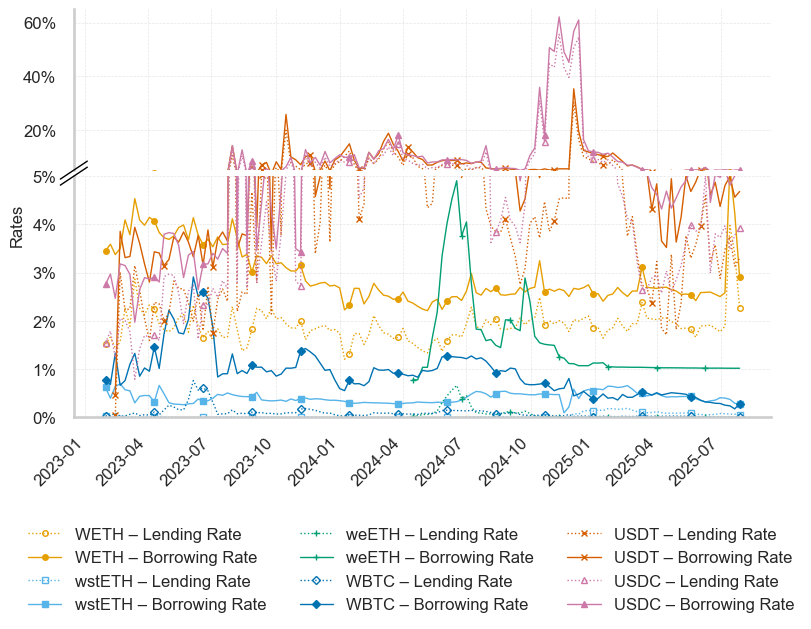}
\includegraphics[width=0.49\linewidth]{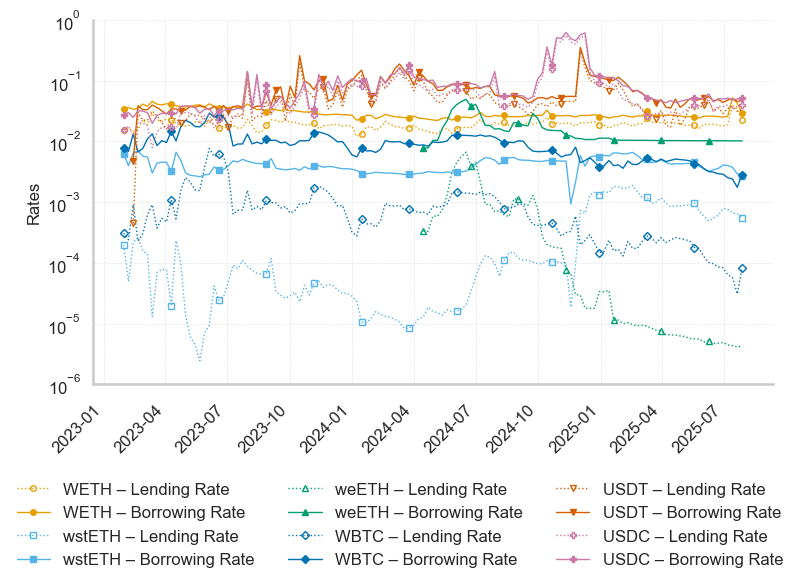}

\includegraphics[width=0.49\linewidth]{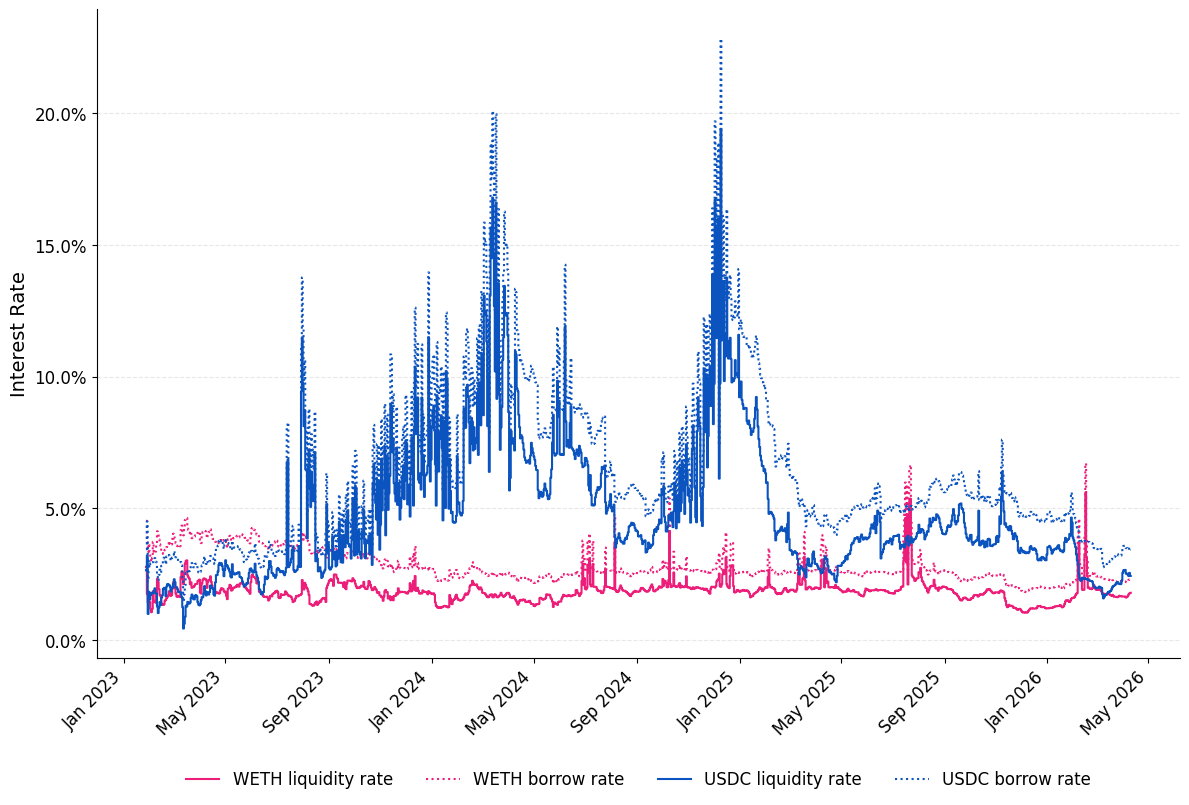}
    \caption{Historical  Lending (Liquidity) and Borrowing rates in Aave protocol.\\
    Top: for a few tokens, for the period January 2023 to August 2025.\\
    Bottom: for USDC and Wrapped ETH, for the period January 2023 to May 2026.\\
    We would like to thank Louis Latournerie for providing us with these plots.}
    \label{fig:rate:Aave}
\end{figure}

\end{document}